\documentstyle[12pt,aasms4]{article}

\def\T1{\ {$T_1$}\ }
\def\CT1{\ {$(C-T_1)$}\ }

\def\gtsim{\ {\raise-0.5ex\HBox{$\buildrel>\over\sim$}}\ }
\def\ltsim{\ {\raise-0.5ex\HBox{$\buildrel<\over\sim$}}\ }

\begin{document}

\title{Statistics of Stellar Populations of Star Clusters and Surrounding\\
Fields in the Outer Disk of the Large Magellanic Cloud}

\author{Jo\~ao  F. C. Santos Jr.}
\affil{Dep. de F\'{\i}sica, ICEx, UFMG,
C.P. 702, 30123-970 Belo Horizonte MG, Brazil}

\author{Andr\'es E. Piatti\altaffilmark{1}, and Juan J. Clari\'a\altaffilmark{1}}
\affil{Observatorio Astron\'omico de C\'ordoba, Laprida 854, 5000, \\ C\'ordoba,
Argentina}

\author{Eduardo Bica}
\affil{Departamento de Astronomia, Instituto de F\'{\i}sica, UFRGS,
        C.P. 15051, 91501-970  \\
Porto Alegre RS, Brazil}

\author{Doug Geisler}
\affil{Universidad de Concepci\'on, Dept. de
F\'{\i}sica, Casilla 160-C, Concepci\'on, Chile}

\author{Horacio Dottori}
\affil{Departamento de Astronomia, Instituto de F\'{\i}sica, UFRGS,
        C.P. 15051, 91501-970  \\
Porto Alegre RS, Brazil}

\slugcomment{Accepted by the Astronomical Journal}

\altaffiltext{1}{ Visiting Astronomer, Cerro Tololo Inter-American
Observatory, which is operated by AURA, Inc., under cooperative
agreement with the NSF.}

\begin{abstract}
A comparative analysis of Washington color-magnitude diagrams 
(CMDs) for 14 star clusters and respective surrounding fields in 
the Large Magellanic Cloud (LMC) outer disk is presented. Each 
CCD frame including field and respective cluster covers an area of 
$185\Box'$. The stellar population sampled is of intermediate age and
metallicity. CMD radial analysis involving star count ratios, 
morphology and integrated light properties are carried out. Luminosity 
functions (LFs) are also presented. 
Two main results are: (i)~Within the range $4<R($kpc$)<8$, the distance
from the LMC center is  well correlated with the average
age in the sense that inner
fields are younger and; (ii)~Beyond $\approx$8kpc the outer fields do 
not show evidence of a significant intermediate-age component in their 
stellar populations, as inferred from red giant clump star counts.
\end{abstract}


\eject

\section{INTRODUCTION}

The stellar populations making up galaxies mix and change as a result of  
dynamical processes and stellar evolution. 
Star clusters can probe the galaxy's history, as they represent a unique
population born   in a star formation burst.
Field stars, as opposed to cluster stars, do not constitute
gravitationally bound systems
and reflect a mix of stellar populations whose 
local nature is revealed more efficiently through individual star
photometry, i.e., through color-magnitude diagrams (CMDs). In spite
of the recent advances in telescope and detector technologies, just a handful
of galaxies can be investigated through this technique, foremost among
them being the Magellanic Clouds. Galaxies beyond the Local Group are not 
yet reachable
at magnitude levels useful for a statistical treatment of their field
stellar populations by using individual star photometry.

Studies of the intermediate-age stellar populations of LMC fields were 
reviewed by Olszewski et al. (1996) and summarized in their Table 2. 
The only study sampling field sizes larger than in present work is the 
photographic one by 
Stryker (1984), but the size of the population observed is much 
smaller. (Note that their Table 2 shows Hodge (1987) observations as sampling
an area of $400\Box'$ although the correct area is $20\Box'$). 
Stryker, by observing fields lying 9\,kpc from the LMC bar, suggested
that a major portion of the stellar content of the LMC is of 
intermediate age.       

 Similarly, Butcher (1977), Hardy et al. (1984), Bertelli et al. (1992), 
Westerlund et al. (1995) and  Vallenari et al. (1996), among others,
observed and analyzed CMDs of field populations in the LMC bar and disk. 
The results obtained from their ground-based observations agree and 
indicate a global burst of star formation to have
occurred in the LMC $\approx$3-4\,Gyr ago.
  
 In particular, Bertelli et al. (1992) observed three fields in the 
LMC located at distances less    than 6$^o$ from the LMC center. 
Their CCD frames 
cover $15\Box'$. They computed ratios of star counts in the CMDs of
fields near the clusters NGC\,1783, NGC\,1866 and NGC\,2155
in order to investigate the star formation history of the LMC. 
Synthetic CMDs were built by using stellar evolutionary models and 
assuming an initial mass function coupled to a  star formation rate 
described by the relative strength of the burst, and the ages of the 
beginning and end of the burst. Their analyses show that a
Salpeter IMF slope is plausible for all fields,  and all are characterized by
having    a burst beginning at a similar age, i.e. 3-4\,Gyr ago. 
After this burst began, the mean star formation rate in the LMC has 
been increased 10 times up to at least 1\,Gyr ago at places as far 
as $\approx5$\,kpc from the LMC center. The star formation in the 
central regions of the LMC is continuing to the present day.      

Observing smaller fields, Westerlund et al. (1995) obtained
different results for two fields $\approx6^o$ apart lying at distances
$\approx4.5^o$ from the LMC center. Their northwest field has a 
dominant stellar component younger (0.3-0.8\,Gyr) than that making up 
their southwest field. Not only fields, but also clusters follow this
trend. Both fields contain a well represented intermediate age 
population (1-3\,Gyr), which led the authors to infer a global 
star formation burst occurred $\approx3$\,Gyr ago.

HST observations, although sampling smaller area fields, produce 
deeper photometry than ground-based observations. Geha et al. (1998),
and references therein,  have investigated three LMC fields at 3$^o$ 
to 4$^o$ from the bar center, one located to the  northeast and two 
to the northwest. They suggested that there has been more star formation
in the past than it was previously believed. Their models, assuming 
a standard IMF slope, indicate a nearly constant star formation rate for 
10\,Gyr, then it increases by a factor of three for the last 2\,Gyr.
On the other hand, Elson et al. (1997) show evidence for a major 
star-forming event to have ocurred $\approx$2 Gyr ago, in agreement with 
ground-based studies.

The studies of LMC intermediate-age star clusters were also reviewed 
by Olszewski et al. (1996). Data for those 
clusters which have CMDs not severely affected by field contamination
were summarized by Geisler et al. (1997). Besides the clusters listed by
Geisler et al. (1997), NGC\,1987 and NGC\,2108 were also 
observed in $UBV$ by Corsi et al. (1994) and NGC\,1806 
has been observed in $JHK$
but the resulting CMD contains only giant stars (Ferraro et al. 1995). 
It is well known that the LMC cluster distribution 
presents a lack of clusters between $\approx$3\,Gyr and $\approx$12\,Gyr, 
the one exception being ESO121SC03, with an age of $\approx$9\,Gyr. 
In a recent study using the HST, Sarajedini (1998) obtained $\approx$4\,Gyr 
for the age of three populous clusters (NGC\,2155, SL\,663 and NGC\,2121), 
claiming not only 
that those clusters fill the age gap if the adopted metallicity
is $[Fe/H]=-1.0$, but also acknowledging the need for a larger sample of
intermediate-age clusters in order to constrain the LMC cluster formation 
history in this age range.  

All of these investigations involve analyses of stellar evolutionary phases
recognizable from their location in the CMD, as well as studies of luminosity
functions. Ratios of star numbers in CMDs were also employed in some
cases.  

In a recent work,
Geisler et al. (1997) employed  the Washington system (Canterna 1976)
in a search for old
LMC star clusters  and narrowed down the uncertainties on the cluster 
formation history of the LMC for ages greater than $\approx1$Gyr. More
recently, Bica et al. (1998) used the same database to
determine that most of their       sample of
clusters and fields, which belong to the LMC outer disk (distance $>4$kpc from 
the LMC bar center), are of intermediate age ($1\leq$t(Gyr)$\leq3$) 
and metallicity ($-1.1\leq$[Fe/H]$\leq-0.4$). Exceptions
are ESO121SC03 (t(Gyr)$\approx9$) and SL\,769 (an inner disk cluster).

This paper presents the third in our series based on the Geisler et al. (1997)
data. The present study involves  an analysis of a sample of 
Washington CMDs of clusters and populous fields located in the LMC 
disk, using star counts as a valuable source of information regarding
stellar population variations throughout that galaxy. A comparison
of cluster and spatially related field stars is another goal which can be 
accomplished through star counts. In some cases, proceeding with
such analysis involves considering  
poor statistics, crowded fields and Galactic star 
contamination, among other effects such  as differential reddening across 
the LMC. In order to count stars in different phases of evolution, a 
set of boxes limiting regions in the CMD was employed making it possible to 
determine  luminosity functions and to evaluate Galactic
star contamination.  

Several  factors make the present work an improvement on previous studies;
the larger field sizes and the variety of  positions and     larger 
distances from the LMC center. Large fields yield better statistics 
for CMD based studies, such as                the radial behavior
of star counts in different evolutionary stages.
Fields spread out in distance give an opportunity to analyse radial 
gradients of stellar  population parameters.

The sample is presented in Sec. 2. The procedure employed
to count stars in the CMDs is described in Sec. 3, where also the 
contamination by Galactic stars is accounted for and luminosity 
functions of cluster and fields are presented.  
Discussions of the morphology of field CMDs,
radial properties of the sample and ratios of star 
counts are given in Sec. 4. Finally, Sec. 5 outlines the conclusions 
and an appendix giving  an analysis of crowding is also included.

\section{The Data}

The observations are detailed in Geisler et al. (1997) and Bica et 
al. (1998). In summary, the 
observations were carried out with the CTIO 0.9m telescope
with the Tek2k \#3 CCD. The scale on the chip is $0.40\arcsec$ 
per pixel, yielding an area $13.6\arcmin\times13.6\arcmin.$
SL769 was observed with the CTIO 4m with the Tek2k \#4 CCD, with similar 
pixel and areal coverage. The filters used for both runs were the
Washington C and Kron-Cousins R filters and the 
observations were calibrated in the  C, T$_1$ Washington system. 
The data were reduced with DAOPHOT II (Stetson 1987) after trimming,
bias subtraction and flat-fielding.
Typically $10\%$ of the detected objects in a frame were discarded as a 
result of adopting criteria aiming at reducing errors. 
More details on the observations,
reductions and calibration procedures were given in Geisler et al. (1997).  

The clusters and fields sample is presented in Table 1. Clusters have
the following catalog designations: SL (Shapley \& Lindsay 1963),
LW (Lyng\aa~ \& Westerlund 1963), ESO (Lauberts 1982), OHSC 
(Olszewski et al. 1988). By columns:
(1) Cluster identification with respective surrounding field; (2) and (3)
equatorial coordinates; (4) and (5) Galactic coordinates; (6) Number of
stars sampled; (7) Distance in degrees from the LMC center, adopted as
the position of the cluster NGC\,1928, $\alpha_{1950}=5^h21^m19^s$ and 
$\delta_{1950}=-69^o31'30''$; (8) Deprojected distance in degrees from 
the LMC center assuming all clusters within the outer disk tilted at
$i\sim45^o$; (9) and (10) Reddening and distance modulus according to 
the relations $E$($C-T_1$)$=1.966E$($B-V$), $A$($T_1$)$=2.62E$($B-V$) and
($m-M$)$_o=18.5$; (11) and (12) Age and metallicity according to 
determinations in Bica et al. (1998).

Relationships between Washington and $UBV$ systems adopted here are those
by Geisler (1996). Namely, two useful equations 
are $V=0.052+T_1+0.256$($C-T_1$) and ($B-V$)$=0.076+0.475$($C-T_1$).  

Except for ESO121-SC03 (an old cluster) and SL\,769 (an inner disk cluster),
the sample of clusters and surrounding fields have intermediate age and 
belong to the LMC outer disk. Their metallicities are in the range
$-1.05\leq$[Fe/H]$\leq-0.35$.
  
\section{Star Count Analysis} 

In the following a detailed description of the procedure adopted to
study the LMC stellar population sampled in the CMDs is presented.

\subsection{Box Definitions}

  By sorting out stars in boxes encompassing fundamental bright  
evolutionary sequences 
in the  CMD, it is possible to count them  and to study absolute 
numbers and their ratios. 
This is especially useful for the fields where crowding is not 
serious and we have basically 
comparable areal extractions in the frames throughout the LMC. 
Boxes were defined  in the CMDs enclosing 14 regions on the 
main sequence (MS) and 17 regions from the subgiant branch (SGB) 
to the giant branch (GB) 
tip, passing by the red giant clump (RGC). Another 2 boxes 
account for Galactic 
contamination. In Figure 1 the boxes are shown together with 
color and magnitude averages of stars within an anullar area 
($500<r(pix)<691$) of the field around SL\,769.   

Luminosity functions (LFs) were derived 
for the MS and the GB with a 0.5 mag. bin in $T_1$ (the boxes' width). The
field CMD of SL769 (youngest in the sample) was used in order to define  
the boxes' length (($C-T_1$) color range)
as well as their
position in the diagram. Such positions were chosen aiming at enclosing 
stars in the main evolutionary sequences. Since the field surrounding 
SL769 contains the youngest population of our sample, the remaining
clusters and fields do not populate boxes in the upper MS.
LFs for all the sample were built with the complete 
set of boxes, independently of the population age, for the sake of 
uniformity. Some clusters have a relatively small number of
stars (see Table 1) and little information can be extracted from their
LFs. On the other hand, most of the fields are populous and 
their LFs are statistically meaningful. 

Star counts within these boxes were performed allowing a comparison 
between the cluster and respective field populations. In order to
get reliable luminosity functions the effect of 
Galactic foreground stars on the star counts was taken into account. 
Crowding was also considered. In the Appendix a theoretical approach 
to evaluate crowding is used and compared to the present photometric 
data. No corrections for incompleteness 
and differential reddening were applied (the latter effect should 
be negligible in all of these outer disk fields).
 
\subsection{Accounting for Galactic Star Contamination}

The extent of field contamination of each cluster CMD was
checked in Bica et al. (1998) by obtaining an equal-area field CMD 
composed of the addition of CMDs derived from 4 different fields, each 
with  an area 1/4 that of the cluster and lying far away from the cluster. 
Such a comparison was an overestimate for the field contamination since 
the photometric limit within the cluster is brighter
than in the field, and more stars are discarded from the cluster because 
of larger photometric errors due to increased crowding. Nevertheless, 
in only 1 outlying cluster was the number of 
stars obtained in the equal-area field significantly higher than 1/4 of 
the stars in the cluster CMD. The typical ratio was only $\sim 10\%$. 
Here a different approach was used to evaluate  field contamination by
foreground Galactic stars on the cluster and field CMDs.

 The CMD of the field surrounding OHSC\,37, the outermost cluster in our
database, was used as a sample of  Galactic field contamination,
since Bica et al. (1998) found no indications of any LMC field stars there.
The Galactic contamination was accounted for in the LMC clusters and fields by 
using two control boxes in the CMDs gauging the number of stars in loci 
clearly off the main evolutionary sequences.
In the field surrounding OHSC\,37, the ratio of    star counts 
within these two Galactic field boxes and the LMC cluster or field 
boxes gives the correction factor used to infer the relative number 
of Galactic field stars on the CMD sample. Since some boxes in the 
field surrounding OHSC\,37 have only a small number of stars, it would 
be best to check such a correction by using more detailed studies.   

 Ratnatunga \& Bahcall (1985, hereafter RB85) estimated the contribution 
of Galactic field stars in various directions including one towards the
LMC. In that direction, considering $E$($B-V$)=0.06, the star density
predicted by the Galaxy models of Bahcall \& Soneira (1980; hereafter
BS80) in the photometric box $0.8<$($B-V$)$<1.3$ and $17<V<19$ is 
0.16~arcmin$^{-2}$. At a fainter bin, $19<V<21$, and the same color range,
the star density is 0.13~arcmin$^{-2}$. 
In order to compare these values
with our Galactic field sample (the field surrounding OHSC\,37), 
it is advisable to note that OHSC\,37 is far from the LMC center,
9$^o$ closer to the Galactic equator and has a higher reddening value 
($E$($B-V$)=0.15). Both factors suggest an increase in star density. 
To measure the star density in the observed field, calibrations 
relating Washington and Johnson photometry were employed (Geisler 1996)
to transform our observed $T_1\times$($C-T_1$) CMD into a $V\times$($B-V$)
one. After performing these transformations,
star counts in the CMD of the field surrounding OHSC\,37 
in the same photometric windows resulted in star densities of 
0.43~arcmin$^{-2}$ and 0.4~arcmin$^{-2}$, respectively, which are 
significantly larger than 
the values given by the model, as expected. A comparison of model star 
densities at different positions and reddenings led us to conclude   
that the discrepancy between the CMD counts and the  RB85 predictions 
is mainly produced by differences in 
Galactic latitude. 
 
The CMD for the field surrounding OHSC\,37 contains 559 stars, which
makes it a statistically good estimator for  Galactic contamination 
towards that direction. The whole sample, however, spreads over    distances 
of    $\approx10^o$ around the LMC center; therefore variations
of the Galactic foreground star density are expected among the LMC
regions analyzed in the present work. There follows a more precise
attempt to see how large is such a variation  and how 
comparable are modeled and observed star densities towards OHSC\,37.

 Given the Galactic latitude
($b$) and longitude ($l$) as well as a $V$ magnitude, the Galaxy models
of  BS80 compute cumulative star counts (for magnitudes brighter than $V$)
towards the indicated direction and allow one to redden the models 
in order to directly compare them with the observations.
Table 2 compares star densities in our observed sample of the
Galactic field with the one predicted by BS80.
For each two-magnitude
bin investigated the observed star density is slightly smaller than
the one predicted by the models, taking into account 
the star density errors quoted by BS80 (25\%). 
It should be noticed that the $V=22\pm1$ bin does not contain a 
statistically complete star sample since there are essentially no 
stars fainter than $T_1=22.5$ ($V\approx22.8$) in the analyzed 
photometry.

The field taken as representative of Galactic contamination
should sample the highest number of foreground stars when compared 
to the remaining 
fields because it is closest to the Galactic plane.
Thus, the correction factor derived from it should be taken as 
an upper limit when applied to the other fields. On the other hand,
Table 2 shows that the star densities used to derive this correction
factor are slightly below those determined theoretically.
The boxes in the CMDs enclosing the main 
evolutionary sequences and nearby regions are a preliminary way to
get rid of stars clearly pertaining to the Galaxy. 
An estimate of the density of Galactic field stars  within the 
CMD boxes was made to get more reliable LFs.
Specifically, the procedure employed to determine the correction factor
was ($i$) to count stars in two boxes clearly off the main evolutionary
sequences present in the CMD of both the field surrounding OHSC\,37 and
the LMC cluster or field of interest;
($ii$) to count stars in a given ``LMC field'' box representing the locus 
of an evolutionary sequence  segment in OHSC\,37;
($iii$) to compute for all field and 
cluster CMDs the number of Galactic foreground stars expected in  
the ``LMC field'' box from the following expression:

\begin{equation}
N_i^{LMC}=N^{LMC} N_i^{OHSC37}/N^{OHSC37}
\end{equation}
where $N$ represents the sum of the number of stars in the two
``Galactic field'' boxes and $N_i$ the number of stars in the i-th
``LMC field'' box; ($iv$) to repeat the
calculation for all ``LMC field'' boxes making possible the 
construction of corrected LFs for each CMD. 

\subsection{Luminosity Functions of Clusters and Fields}

Figure 2 presents the LFs of clusters and fields without (histograms)
and with (continuous curves) the correction for Galactic
foreground. Left panels are MS LFs and right panels are GB LFs.
No strong effect on the LFs morphology is seen for any cluster or 
field as a consequence of the Galactic foreground. 
Fig. 2(a) shows in the top panel the LF for the field used as a sample
of the Galactic foreground. The field surrounding ESO121-SC03 is the 
only one for which there are significant differences between the 
corrected and observed LF as can be seen from Fig. 2(b) lower 
panel. Fig. 2(c) presents the most populous fields
LFs. Clusters and fields in general agree regarding the RGC location.
A conspicuous old  SGB ($2<M_{T1}<3$)  and its 
corresponding ascending GB ($1<M_{T1}<2$) are visible only in 
the fields LFs. 
Although a small number of stars were present in the MS of most
of the clusters, three of them (SL\,388, SL\,509 and SL\,842) seem to 
have a shallower MS LF as compared to the corresponding surrounding 
field. Differences were found between cluster and respective field 
in the MS LF peak, which could be due not only to crowding but, more likely, 
age effects. ESO121-SC03 and SL\,451 are clearly clusters 
which have the maximum of the MS LF at the same magnitude as their
respective fields.

\section{Discussion}
\subsection{Field CMDs Morphology}

Is there a radial variation of any astrophysical property in    
each field considered individually? In order to answer such a question,
the radial variation of the mean locus
of the main evolutionary sequences in the CMDs for the more populous 
fields was investigated.
The set of boxes was used to define a mean CMD locus at different
distances from the field center (coincident with the corresponding
cluster center). The mean CMD locus should be
understood as the position in the CMD given   by the flux-averaged 
$T_1$ magnitude and the mean ($C-T_1$) color of the stars within
each box. Fig. 3 shows the CMD sequences for four equal area
annular fields surrounding SL\,769 and for three of these annular 
fields surrounding SL\,388 and SL\,509. The mean
CMD sequences for the entire field ($r>150$pix) are also shown. 
It is clear       
that there are no important radial changes in the
CMD morphology of the four most populous fields in our sample
and, as a consequence, any  possible color gradient
of the central cluster no longer  survives at those distances.
The scatter for
the brightest ``mean stars'' on the MS and GB is a consequence of
the small number of stars in the corresponding boxes.  It
allows one to estimate at which magnitude the stochastic nature
of the mass function becomes significant and how it depends on the
size of the population. Table 3 shows the star number density, the 
integrated absolute $M_v$ magnitude (as computed using the 
transformation relation in Geisler (1996)), the integrated ($C-T_1$)$_o$ 
color, and the integrated absolute 
$M_{T1}$ magnitude for each annular field and the complete one allowing 
an estimate of how significant the stochastic effects are on the
integrated light of those populous fields. Specifically, the spread
in ($C-T_1$)$_o$ is 0.1 for SL\,769, 0.12 for SL\,509 and 0.25 for
SL\,388, which are in order of decreasing number of stars. By enlarging
the sample of populous fields it will be possible to better quantify
these effects on purely observational grounds  by means of a 
statistically meaningful relation between the color changes and the 
population size.
As such, this procedure will provide constraints for stellar population
studies of composite systems for which the integrated light is the only
accessible observable.

\subsection{Radial Properties of the Sample}

Is there a radial variation of any astrophysical property along 
the LMC disk as sampled from the whole set of fields? This question
was addressed by using the data in Tables 1 and 3. The density of stars 
in specific evolutionary phases for the fields was also analyzed as a 
function of distance. Ages for the  youngest field population  
were estimated and their radial behavior investigated.
  
\subsubsection{Integrated light and density}

Figure 4  shows that the  behavior of  integrated properties
with distance for clusters (open triangles) and fields (filled 
circles) is not, in general, the same. Although the poor statistics
of some clusters may affect their apparent distribution, the fields are
in general sufficiently populous to make a comparative interpretation 
for them worth doing.

The projected stellar density of the fields decreases more abruptly 
towards the outer LMC regions than that of the clusters (Fig. 4(a)). 
A possible 
cause for this is suggested in the following. Both field stars 
and cluster stars are under the influence of the
same LMC gravitational field at a given distance. Cluster stars,
though, are bound due to their self-gravitation. External
gravitational fields can disrupt clusters in the inner
regions of the LMC. So, moderately dense clusters survive, 
in principle, at any distance in the range $4-13$kpc 
from the LMC center. The slight decrease of cluster star density
with distance seen in Fig. 4 appears to be a consequence of a
lack of low density (less bound) clusters for the inner 
regions, which may have been destroyed by the tidal forces 
of the LMC. On the other hand, the field stars 
are distributed according to the LMC gravitational field, from
which the steeper decrease of density as a function of distance is 
expected: the closer to the LMC center, the denser the field is.

The ($C-T_1$)$_o$ color seems to be redder
for outer fields when compared to inner ones (Fig. 4(b)). The same 
happens for
clusters but the scatter is much larger. The surface $M_{T1}$ (magnitude
obtained from the sum of the individual stars flux per total area observed)
for the fields decreases, on average, with the distance from the LMC 
bar, an expected result (Fig. 4(c)). Although a much more scattered and flat
relationship exists for the clusters, in this case a selection effect is 
in play, since low surface brightness clusters can only
be seen far from the LMC bar. Regarding this selection effect, it
is interesting to note that ESO121-SC03 would be very difficult to distinguish
if it were at a distance $R\leq7^o$ from the LMC center. 

According to Fig. 4(d), the total
($C-T_1$)$_o$ obtained from the CMD sum of the flux of individual stars 
has a spread much larger in the clusters ($\Delta$($C-T_1$)$_o\approx1.5$) 
than in the fields ($\Delta$($C-T_1$)$_o\approx0.5$). This result
is consistent with that obtained by Bica et al. (1998), namely, 
the metallicity range of the intermediate-age LMC clusters is larger
than that of their respective fields.

\subsubsection{Density of field clump and MS stars} 

 Two of the boxes used for building the LFs were
employed as representative of the RGC ($-0.35<M_{T1}<0.65$ and 
$0.65<$($C-T_1$)$_o<2.0$). These two boxes enclose the dual clump
structure present in the fields of SL\,388 and SL\,509. The dual 
clump seen in the CMDs of these fields
can be caused by two stellar populations situated at different 
distances (the fainter clump being the signature of a similar 
age population or of an older population presenting a red horizontal 
branch) or by an evolutionary effect in a single population. In the
former case, the presence of a dwarf galaxy located at the distance
of the SMC is a possibility (Bica et al. 1998).
Concerning the last case, Girardi et al. (1998) have predicted 
theoretically the developing of a secondary clump structure in a
$M_I$$\times$($V-I$) diagram of metal-rich populations as a result 
from the fact that stars slightly heavier than the maximum mass
needed to acquire degenerate He cores define a fainter
clump 0.3 mag. below the red primary clump. Presently, it is 
premature to apply one of these explanations to our sample. 
Ideal targets aiming at clarifying such issue are
LMC populous fields, for which further observational   
investigation of the dual clump structure is under way. In the present 
work the  average age of stellar populations is considered and, 
therefore, the dual clump structure can introduce some bias in 
the star counts of the fields surrounding SL\,388 and SL\,509.

The projected density of stars in the RGC phase for each field was plotted
as a function of the deprojected distance $R$ in Fig. 5(a). Figs. 5(b-l)
show the projected density of stars in 11 segments of the MS.
These segments are within the magnitude range 
$-2.35<M_{T1}<3.15$ sampling 0.5 mag. bins each. 
All densities were corrected for Galactic contamination.
The drop in the density
towards the outer regions of the LMC disk can immediately be
noted when one takes any CMD segment and follows it from small to large 
$R$ in Fig. 5. The density of RGC stars increases inward from 
8$^o$ ($\approx$8kpc). At larger distances, just a few  clump stars  are 
seen for the sampled fields. Indeed, in virtually all diagrams, little or no
evidence for a substantial LMC field population is found beyond $\sim10^o$,
in good agreement with the results of Bica et al. (1998).

\subsubsection{The youngest field populations}

The fields are a mix of populations of different ages. One can associate
different parameters in the CMD of the fields to different ages. The magnitude
difference between the most populous turnoff and the RGC is one such parameter 
to derive age for the associated population as made by
Bica et al. (1998). In the present study an age derived from the 
brightest field turnoff, revealing the age of the most recent burst of star
formation, was determined. It relies on the turnoff $R$ magnitude obtained
from Z=0.006 classical isochrones by VandenBerg (1985). Errors were
estimated based on the spread in $T_1$ of the brighter MS stars. Fig. 6
shows this age as a function of the density of clump stars (a), the 
density of MS stars within the magnitude range $1.65<M_{T1}<2.15$ (b), 
the galactocentric distance (c), and the age derived for the richest 
turnoff (d).

Since the fields contain composite populations it should
be recalled that the ages analyzed in the following are averaged ones.
There is a clear correlation between
the age of the youngest burst and the clump star density: in the
range $8<\log{(age)}<9.5$, the higher the density, the younger is the 
most recent burst (Fig. 6(a)). This trend is not as easily discerned 
for the density of MS stars (Fig. 6(b)) but is still present.
The farther the field from the LMC center, the
older its youngest population (Fig. 6(c)). The age derived for the
youngest field population does not appear to correlate with the age
derived from the richest turnoff (Fig. 6(d)). 
In summary, if these are general results valid throughout the LMC disk,
the most recent burst of star formation for a field 
can be dated from its distance from the LMC center (if it is
lower than $\approx$8kpc) and/or its density of clump stars.

Inner fields have younger stellar population components than outer
ones and, as a consequence, the density of clump stars changes 
accordingly. Considering that field and cluster populations have 
similar properties, these results are in agreement with those in Bica et 
al. (1996), namely, that the younger star clusters and associations are
located preferentially in the LMC inner regions while the older clusters
are distributed all over the LMC. 
  
\subsection{Ratios of Star Counts as a Function of Astrophysical Parameters}
   
 Bertelli et al. (1992) computed appropriate ratios of star counts for 
stellar fields in the LMC providing information on the age and 
intensity of star formation bursts.
In the present work the ratio between the number of RGC stars and 
the number of MS stars within a given magnitude range was investigated 
as a function of $R$, $[Fe/H]$ and $age$ for the fields according to 
the values computed in Bica et al. (1998). A segment of the MS, judged 
to be the  midpoint between the turnoff and the observational limit
($2.15<M_{T1}<2.65$ and $-0.35<$($C-T_1$)$_o<0.9$), was used. 
The number of
stars in the field CMDs was counted and the ratio N(RGC)/N(MS) computed
before and after the correction for  Galactic contamination. 
Fig. 7 shows the relations of the corrected ratio with astrophysical
parameters as well as with the uncorrected ratio, the latter
giving an estimate of how the Galactic field foreground stars are
affecting the counts in the two boxes used to compute that ratio.
In Fig. 7(b) this relation is shown and one can see
that the corrected ratio yields lower values than the observed one,
which means that contamination is higher on the RGC than on the MS.
The point strongly affected by the correction is the field surrounding
ESO121SC03. Essentially no correlation can be seen between the corrected 
ratio and $R$, $age$ or $[Fe/H]$, unless  one considers only
the most populous fields (circled squares in Fig. 7(a)). In fact, in these
cases there appears to be a tendency towards a
correlation with metallicity in the sense that the higher the ratio 
the lower the metallicity. The deprojected distance
is uncorrelated with N(RGC)/N(MS) according to Fig. 7(c).
Since the sample spans a narrow range in
age (except for ESO121SC03), Fig. 7(d) evidences a large 
range of the ratio (0.2-0.55) for a single age. 
The outer fields surrounding SL126 and  SL842 
are labeled in Figs. 7(c) and 7(d). They present, together with 
ESO121SC03, the oldest average ages
of the sampled fields. The very low ratio for ESO121SC03 is probably an
artifact of small number statistics.

\section{Summary and Concluding Remarks} 

 With the goal of studying statistically the stellar properties of
14 clusters and surrounding
fields in the LMC, boxes were defined in the CMD of these objects.
The number of stars within  each of these boxes, 
which sample the main evolutionary sequences (RGB, clump, turnoff),
were used to build and compare the luminosity functions of clusters
and respective fields. The effect of Galactic
star contamination on these luminosity functions was also estimated. 
In view of large CCD 
areal coverage ($13.6'\times13.6'$) yielding statistically complete CMDs, it  
was possible to look for LMC stellar content differences  from cluster to 
respective field and from field to field.

For each field CMD, mean evolutionary sequences 
were computed based on the mean magnitude and color of the stars 
within each box. The more populous fields allowed exploration
of a possible radial change of the CMD morphology. 
The CMD structure of the inner populous field around SL\,769 allowed us
to estimate an integrated color and magnitude and their sensitivity to a
changeable small number of bright stars on the giant branch and main 
sequence. Fluctuations in color and magnitude were
found and quantified for the MS turnoff and GB tip. This analysis provides
a useful constraint for stellar population studies of composite 
stellar systems for which only integrated light is available.

Besides CMD morphology, star counts and ratios in 
well--defined evolutionary sequences were carried out.
No correlation was found between the  ratio of clump stars to
MS stars within a given magnitude range and the 
distances of the fields from the LMC center. 
On the other hand, star 
counts in specific evolutionary phases (RGC and MS) did show a 
correlation with galactocentric distance reflecting the coupled effect 
of decreasing density and increasing age towards the outer fields 
investigated.     

Luminosity functions for clusters and fields were presented.
Regarding the most populous field LFs, SL 388 and 509 seem to have a 
shallower main sequence LF slope than its associated field.  
Since the observed
clusters and fields are located at various distances from the LMC center,
reaching as far as $\approx13$kpc, possible spatial variations of the 
LMC stellar population were assessed. There is some indication of a 
($C-T_1$) radial gradient at least from 4 to 8($^o$) from the LMC center
in the sense that  the farther the field the redder it is. 

An enlargement of the sample of fields in the LMC at various
distances from the bar will certainly clarify correlations that
could only be suggested with the present 14 fields, such as      
the increase of the ($C-T_1$) color with increasing distance.

\bigskip
This research is supported in part by NASA through grant 
No. GO-06810.01-95A (to DG) from the Space
Telescope Science Institute, which is operated by the Association 
of Universities for Research in Astronomy, Inc., under NASA 
contract NAS5-26555. This work was partially
supported by the Brazilian institutions CNPq and FINEP, 
the Argentine institutions
CONICET and CONICOR, and the VITAE and Antorcha foundations.
We acknowledge an anonymous referee for interesting remarks.

\bigskip
\bigskip
\begin{appendix}
\centerline{APPENDIX: Accounting For Crowding} 

Renzini (1998) presented a theoretical framework useful for the
analysis of CCD photometry whenever individual stars are to be
resolved as is the case in CMD studies. 
The total luminosity of the stellar population sampled in the 
CCD frame is related to the number of stars in each evolutionary 
stage. His study also allows one to see at which magnitude 
crowding effects become significant, after which the 
probability of a pixel or resolution element to contain a 
blending of stars is extremely high. At this magnitude level the 
observed luminosity function would be severely influenced 
since two or more faint stars would be mistaken as a single 
brighter star. In order to check
the correspondent threshold magnitude, the total bolometric
luminosity per  pixel was computed for the cluster sample, 
since the fields are not expected to be affected in a 
significant way by crowding. 
The total bolometric luminosity for each cluster was estimated 
according to:

\begin{equation}
L_T=k L_v=k 10^{-0.4(M_v-M_{v\odot})}
\end{equation}

where $k\sim1.5$ is the ratio between bolometric and $V$ band 
luminosities for intermediate age populations (Renzini 1998); 

~~~~~~~~~$M_{v\odot}=4.82$ is the absolute $V$ magnitude of the sun;

~~~~~~~~~and $M_v$ is the integrated absolute $V$ magnitude of 
the cluster.

 In order to compute $L_T$, a transformation between Washington 
magnitudes and $V$ from Geisler (1996) was combined with
$A$($T_1$)$=2.62E$($B-V$) and $E$($C-T_1$)$=1.966E$($B-V$). 
If $A(V)=3.12E(B-V)$ is adopted the following relation results

\begin{equation}
M_v=0.052 + 0.256 (C-T_1)_o + M_{T1} + 0.0017 E(C-T1)
\end{equation}

where ($C-T_1$)$_o$ and $M_{T1}$ are the integrated color and
magnitude of the clusters which can be obtained directly from
the observed CMDs. Since the last term in equation (2) is 
negligible, $L_T$ results

\begin{equation}
L_T=k 10^{-0.4(0.052 + 0.256 (C-T_1)_o + M_{T1}-M_{v\odot})}
\end{equation}

Then the total bolometric luminosity per pixel is given by

\begin{equation}
l_T=L_T/N_{pix}
\end{equation}

where $N_{pix}$ corresponds to the sampled area in pixels.

In this way, we   obtain the total light per pixel, which is
to be compared with the light of a single star. 
Relatively small magnitude errors can only be obtained for 
stars brighter than $l_T$. For example, if the magnitude corresponding
to $l_T$ is faint, only stars which are even brighter should 
be considered in a CMD analysis. Then,  
the threshold apparent magnitude at which crowding becomes 
important is derived from (4) by substituting $L_T$ from (3), i. e.
\begin{equation}
T_1=M_{v\odot}-0.052+(T_1-M_{T1})-0.256[(C-T_1)-E(C-T_1)]-2.5\log{\frac{l_T}{k}}
\end{equation}  

where every quantity refers to a pixel.

In practice two different observables were employed to get
$L_T$: ($i$) 
the cluster photoelectric integrated $V$ magnitude (Bica et al. 1996)  
through equation (1) and ($ii$)  the number of stars used to build the 
CMD through equation (3). In order to compute $N_{pix}$ in the former
case, an equivalent number of pixels was determined from the area
observed through the diaphragm using the same scale as the CCD
observations. 
These two approaches give different results since CMDs require 
individual stars to be  selected on the basis of a  well-defined point 
spread function and crowding affects it, therefore preventing a
complete sampling of the observed cluster stars. On the other hand,
integrated light measurements are not affected by crowding. As a
consequence, the total $V$ flux computed from the number of stars 
in the CMDs is  smaller than the one computed from the integrated 
$V$ magnitude if the observed area is the same. 

Table 4 shows the results for the cluster sample. Since the threshold
$T_1$ magnitude depends on ($C-T_1$),  it was  computed for a red star
(($C-T_1$)$=3$) and a blue one (($C-T_1$)$=0$) from equation (5). 
The fact that all clusters
present threshold $T_1$ fainter than that for the dimmest stars 
observed in the CMDs demonstrates that the selection criteria used
to get good quality photometry already discarded stars in crowded 
regions. Figure 8 presents cluster total aparent magnitudes ($V$ or 
$T_1$) from integrated photometry (open triangles) and addition of 
CMD stars flux (filled circles) as
a function of the threshold magnitude for red stars and blue ones.
$\Sigma{T_1}$ is the total cluster
aparent magnitude in the $T_1$ band and $\Sigma{V}$ in the $V$ band. 
Observed blue $T_1$ is the 
fainter magnitude measured for a blue star in the cluster CMDs.
In order to compare exclusively CMD based properties the lower left 
panel does not show cluster total aparent magnitudes from integrated 
photometry.
The upper panels show that the brighter the cluster, the brighter
the threshold magnitude is, this trend being more evident for the 
total $V$ from integrated light. No correlation was found between the
CMD $V$ flux and the observed threshold for blue stars in our CMD sample
as one can see in the lower left panel.
In general, integrated photometry gives brighter $V$ fluxes than 
CMD sums, as expected. 
\end{appendix}

\newpage

\begin{figure}
\figcaption{The 31 CMD boxes enclosing the main evolutionary sequences
presented by intermediate-age stellar populations plus the 2 boxes used
for measuring the Galactic contamination (empty wider   boxes). Dots 
mark the
flux averaged $T_1$ magnitude and mean color of the stars, within each
box, for an annular field surrounding SL\,769 ($500<r(pix)<691$).}
\end{figure}

\begin{figure}
\figcaption{(a) Histogram of the luminosity function for cluster and
respective fields separated into the MS LF (left panel) and the GB LF
(right panel). Solid      lines represent cluster LFs and dashed lines
field LFs. LFs corrected for Galactic contamination are shown as 
continuous curves. The field shown in the upper panel is the one used
as a sample of the Galactic field. (b) Same as (a) for different
clusters and fields. (c) Same as (a) for the most populous fields.}
\end{figure}

\begin{figure}
\figcaption{The CMD mean sequences derived using the boxes for
three populous fields. Each symbol defines sequences corresponding
to a ring centered on the associated cluster. The filled symbol shows 
the mean sequence for the whole field.}
\end{figure}

\begin{figure}
\figcaption{Clusters (triangles) and fields (filled circles) integrated
magnitudes are plotted as a function of integrated ($C-T_1$)$_o$. All 
integrated indices are also shown against distance from the
LMC center (see text for details).}
\end{figure}

\begin{figure}
\figcaption{The density of field  clump stars and of stars in 
11 MS CMD regions against galactocentric distance. The sequence runs 
from brighter (MS1) to fainter (MS11) regions.}
\end{figure}

\begin{figure}
\figcaption{The age of the youngest stellar population in each field is
shown as a function of the (a)~RGC stars density, (b)~lower MS stars
density, (c)~galactocentric distance, and (d)~age of the most populous
turnoff.}
\end{figure}

\begin{figure}
\figcaption{The ratio of the number of RGC stars to the number of MS
stars for the fields within a bin of $\Delta$$M_{T1}=0.5$, corrected 
for Galactic contamination,  is shown as a function of (a) metallicity,
where the most populous fields are circled,
(b) the uncorrected ratio, (c) the deprojected distance, and (d) age.}
\end{figure}

\begin{figure}
\figcaption{Total cluster aparent magnitudes obtained from 
observed integrated
photometry (triangles) and from addition of CMD star fluxes (filled
circle) are presented as a function of the threshold magnitude, 
fainter than which
crowding prevents good photometry. $\Sigma{T_1}$ is the total cluster
aparent magnitude in the $T_1$ band and $\Sigma{V}$ in the $V$ band. 
Red and blue $T_1$
correspond to the threshold magnitude, as computed from equation (5) and
listed in Table 4, for a ($C-T_1$)$=3$ star and
a ($C-T_1$)$=0$ star, respectively. Observed blue $T_1$ is the 
threshold magnitude measured for a blue star in the cluster CMDs.
No cluster observed integrated photometry is shown
in the lower left panel in order to see more clearly how the CMD total 
flux behaves with an observed threshold magnitude.}
\end{figure}


\begin{references}

Bahcall J.N., Soneira R.M. 1980, ApJS 44, 73

Bertelli G., Mateo M., Chiosi C., Bressan A. 1992, ApJ 388, 400

Bica E., Clari\'a J.J., Dottori H., Santos Jr. J.F.C., 
Piatti, A.E. 1996, ApJS 102, 57

Bica E., Geisler D., Dottori H., Clari\'a J.J., Piatti A.E.,
Santos Jr. J.F.C. 1998, AJ 116, 723

Butcher H. 1977, ApJ 216, 327

Canterna R. 1976, AJ 71, 228

Corsi C.E., Buonanno R., Fusi Pecci F., Ferraro F.R., Testa V.,
Greggio L. 1994, MNRAS ~~~~271, 385

Elson R.A.W., Gilmore G.F. \& Santiago B.X. 1997, MNRAS 289, 157

Ferraro F.R., Fusi Pecci F., Testa V., Greggio L., Corsi C.E.,
Buonanno R., Terndrup D.M., ~~~~Zinnecker H. 1995, MNRAS 272, 391

Geha M. et al. 1998, AJ 115, 1045 
(((JS) AJ accepts 'et al.' for a long list of authors))

Geisler D. 1996, AJ 111, 480

Geisler D., Bica E., Dottori H., Clari\'a J.J., Piatti A.E. \& 
Santos J.F.C. Jr. 1997, AJ 114, ~~~~1920

Girardi L., Groenewegen M. A. T., Weiss A., Salaris M. 1998, MNRAS 301,
149

Hardy E., Buonanno R., Corsi C. E., Janes K. A., Schommer R. A. 1984,
ApJ 278, 592

Hodge P.W. 1987, PASP 99, 730

Lauberts A. 1982, The ESO/Uppsala Survey of the ESO (B) Atlas 
(Munich:ESO)

Lyng\aa~ G., Westerlund B.E. 1963, MNRAS 127, 31

Olszewski E.W., Suntzeff N.B., Mateo M. 1996, ARAA 34, 511

Olszewski E.W., Harris H.C., Schommer R.A., Canterna R.W. 1988,
AJ 95, 84

Ratnatunga K.U., Bahcall J.N. 1985, ApJS 59, 63

Renzini A. 1998, AJ 115, 2459

Sarajedini A. 1998, AJ 116, 738

Shapley H., Lindsay E.M. 1963, Irish AJ 6, 74

Stetson P. B. 1987, PASP 99, 191

Stryker L.L. 1984, ApJS 55, 127

Vallenari A., Chiosi C., Bertelli G., Ortolani S. 1996, 
AA 309, 358

VandenBerg D.A. 1985, ApJS 58, 711

Westerlund B.E., Linde P., Lyng\aa~ G. 1995, AA 298, 39

\end{references}
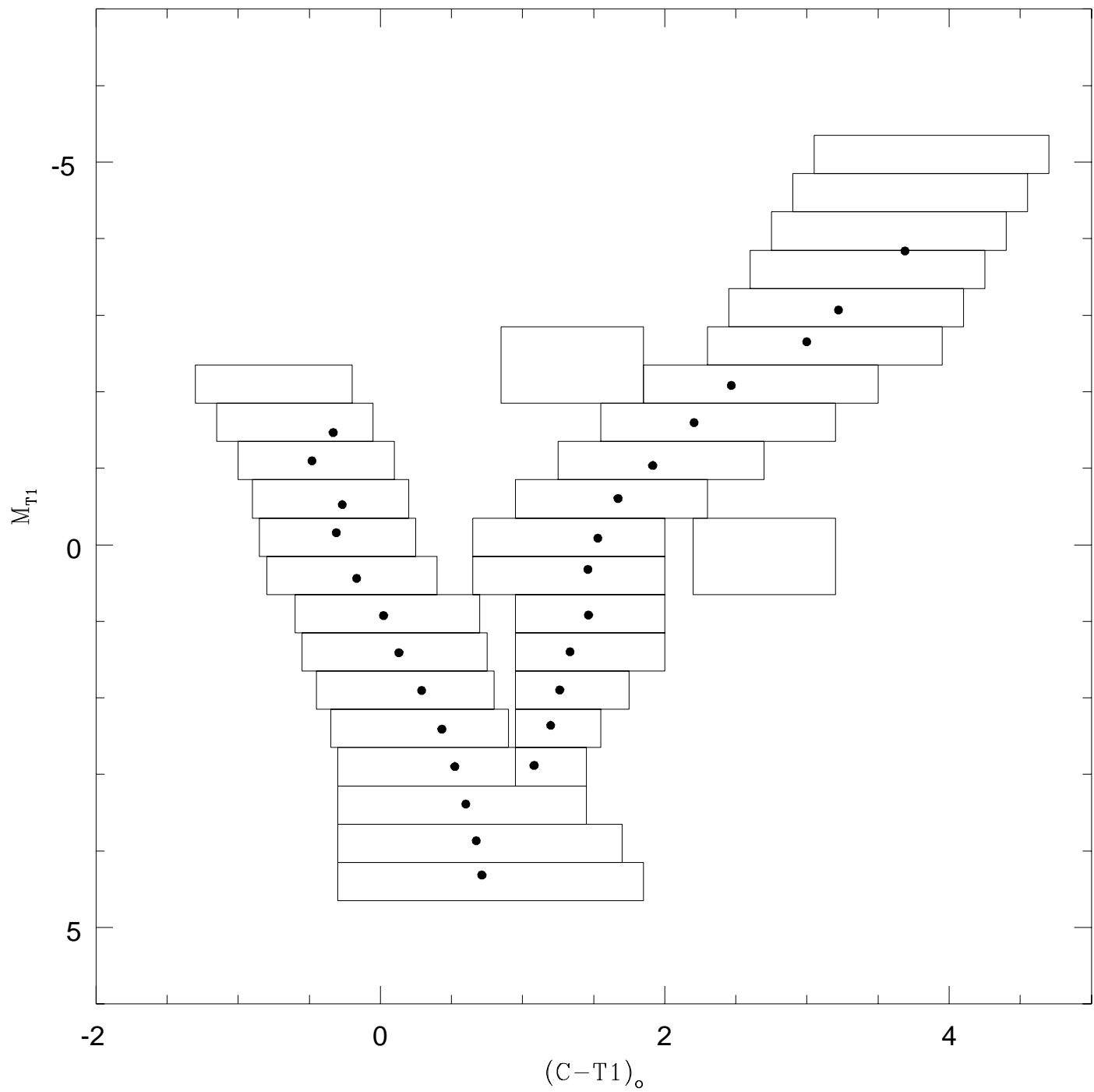
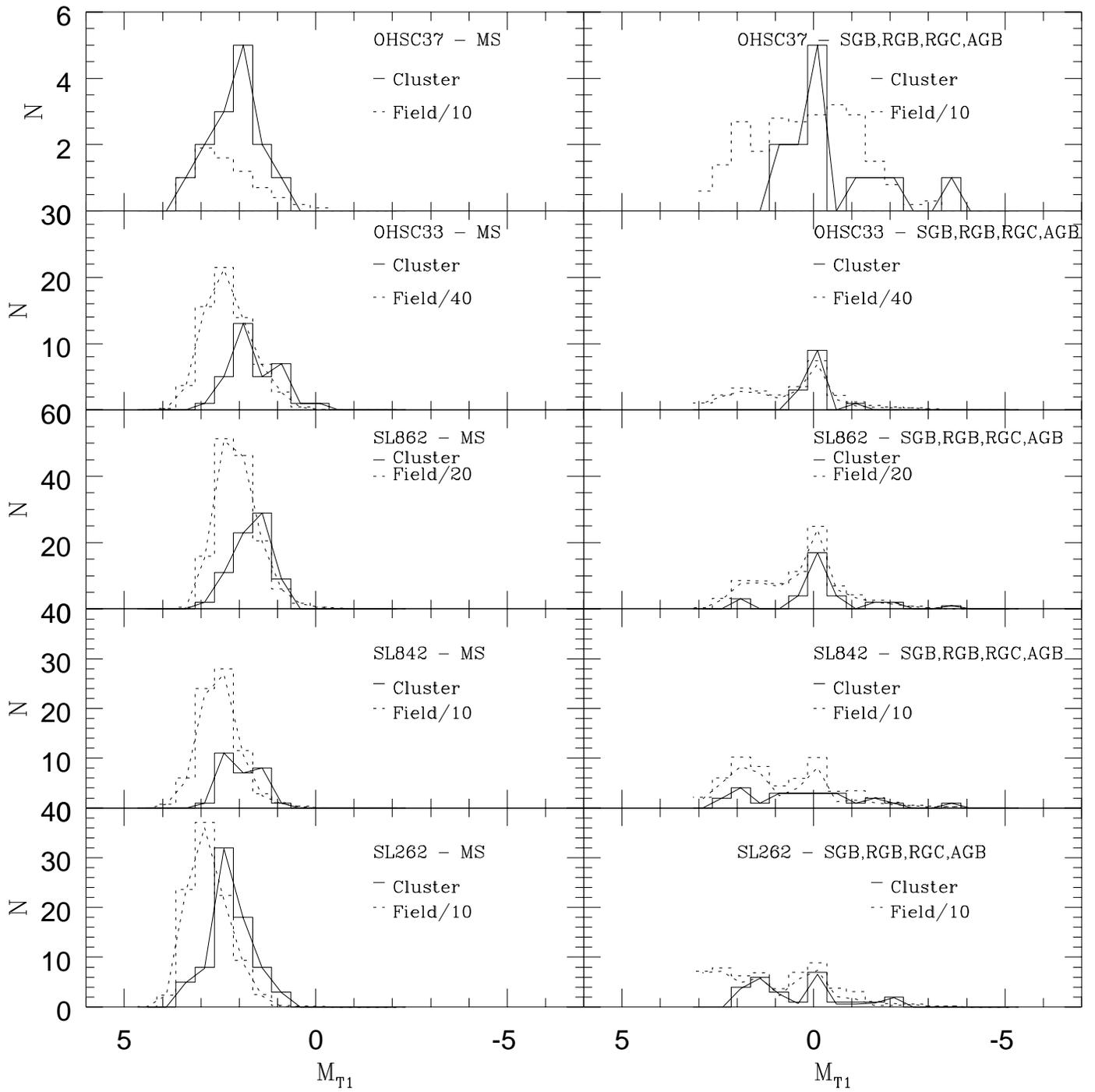
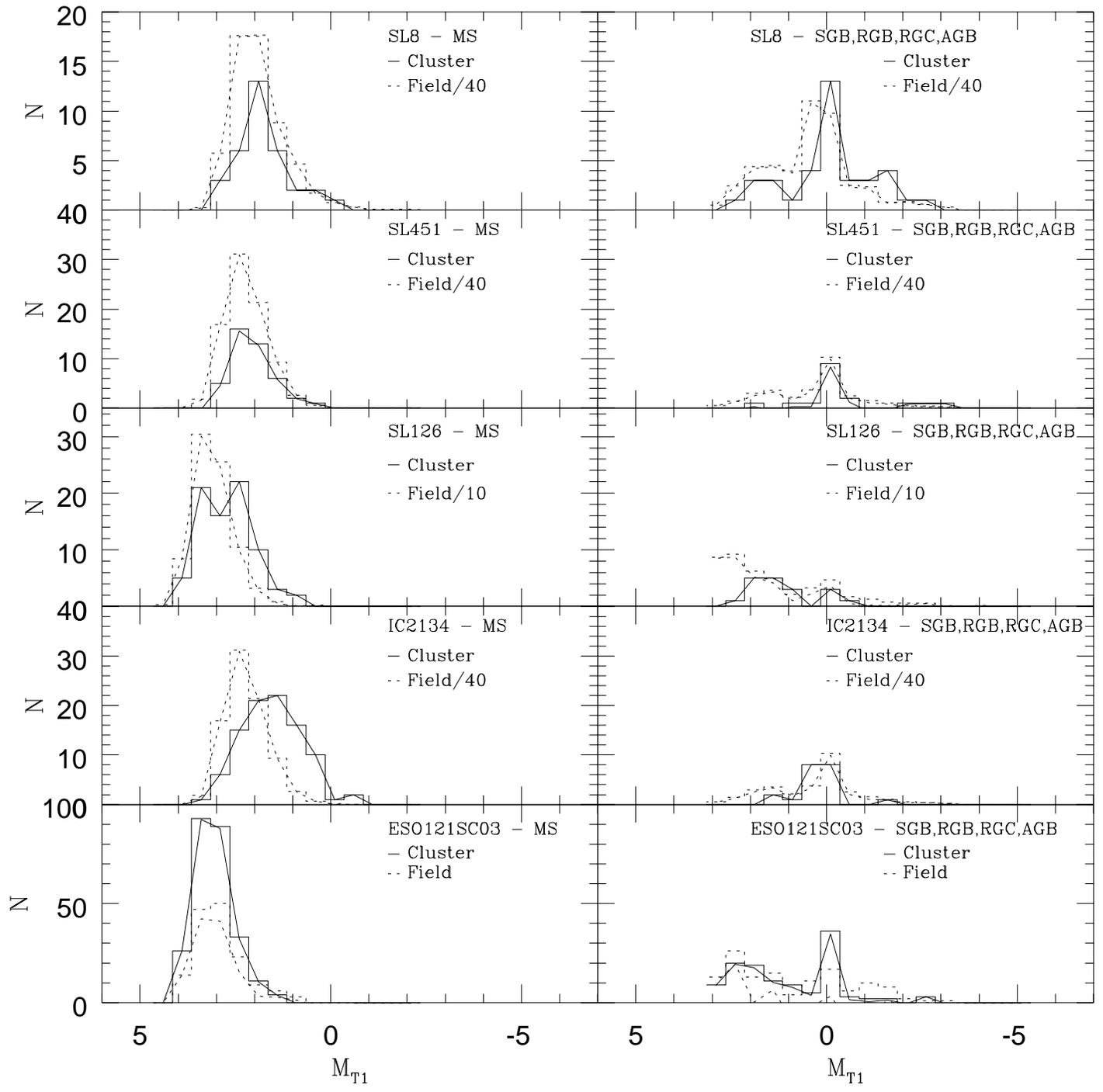
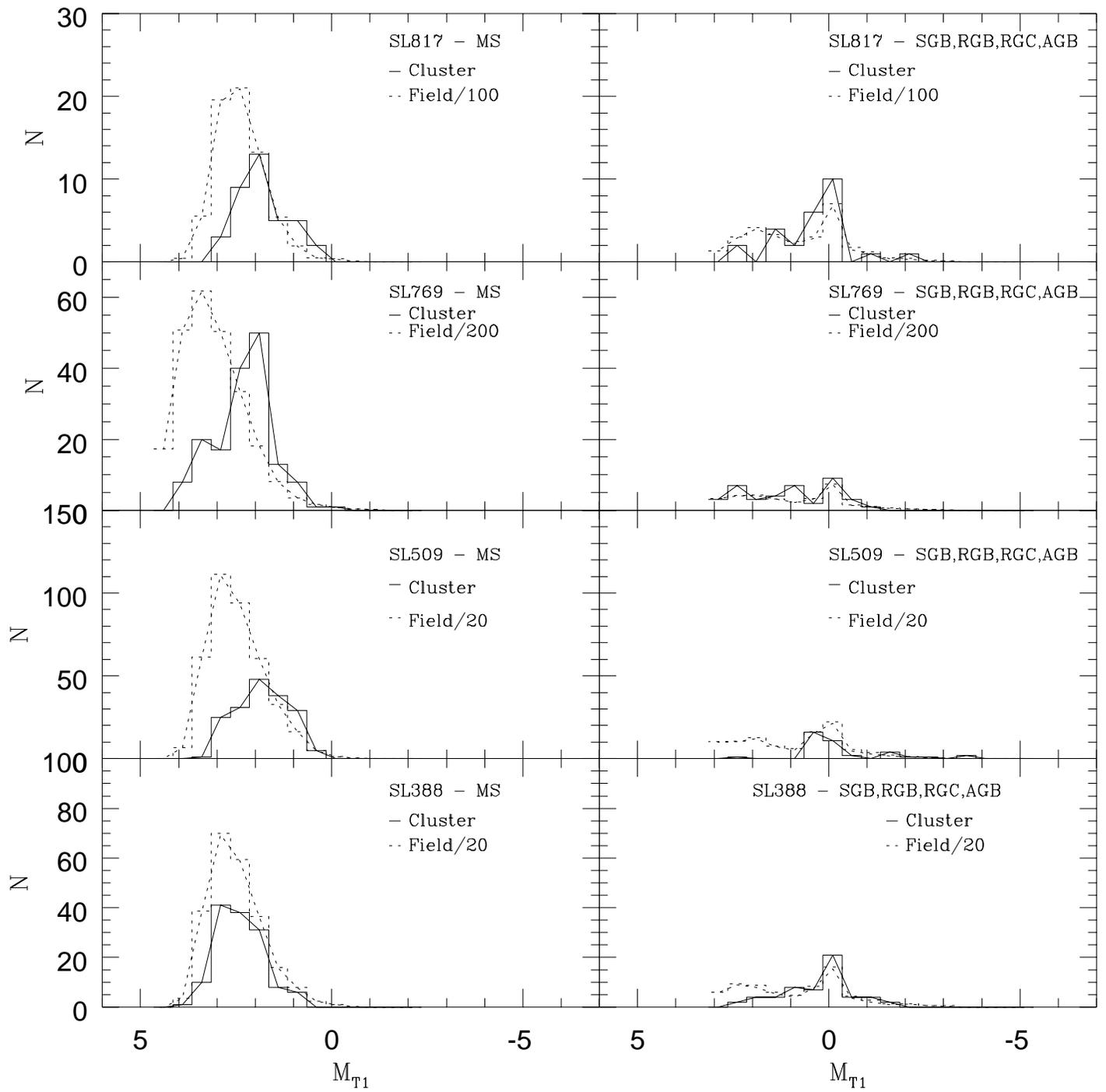
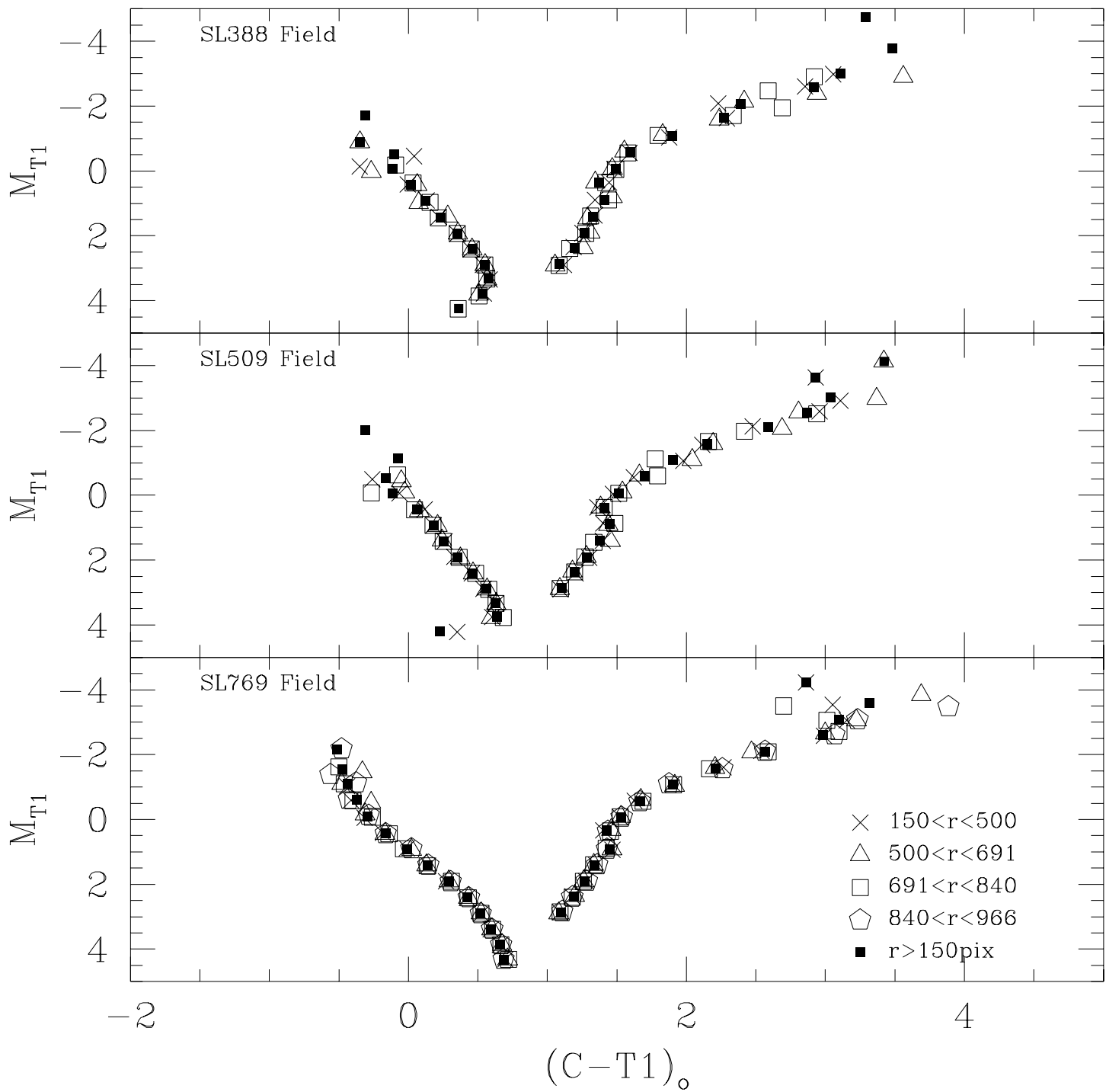
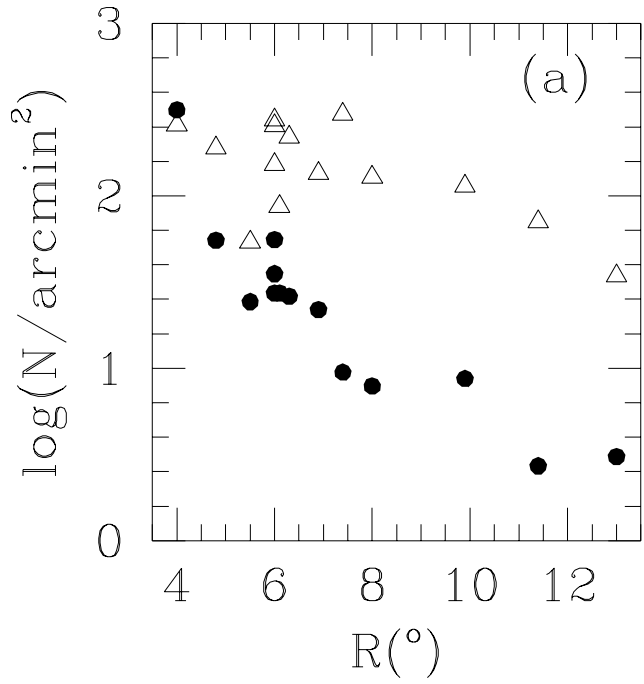
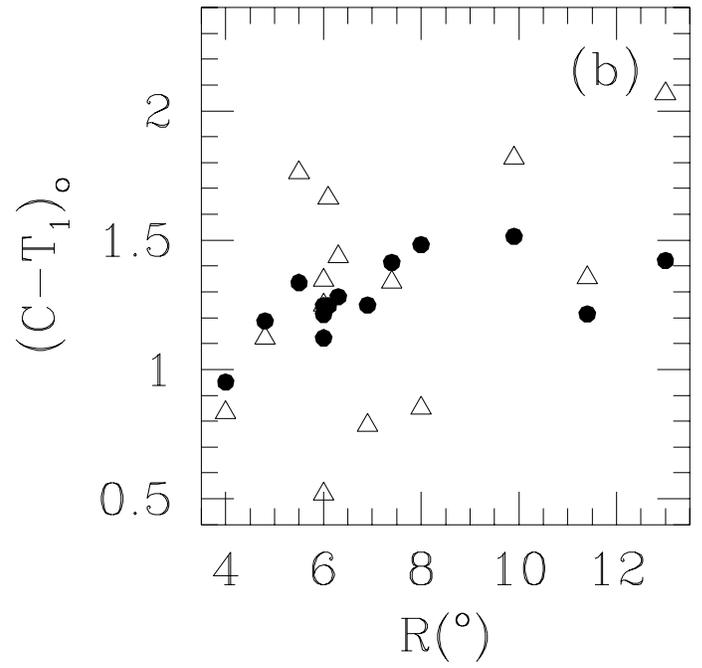
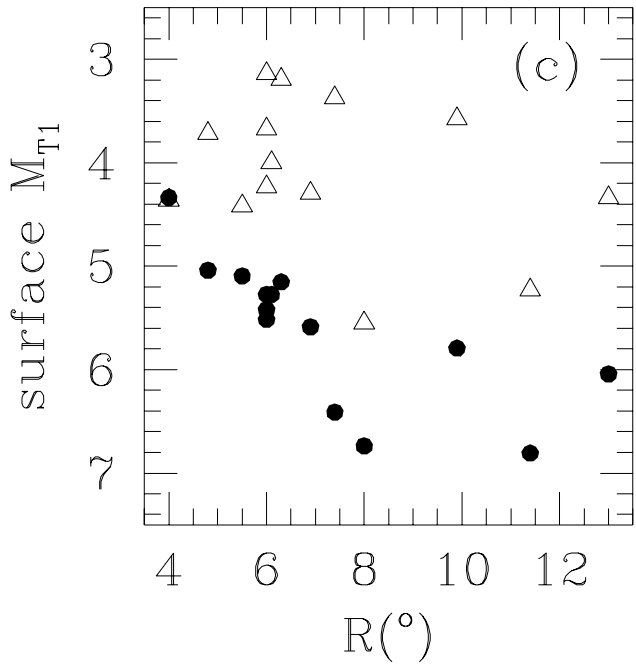
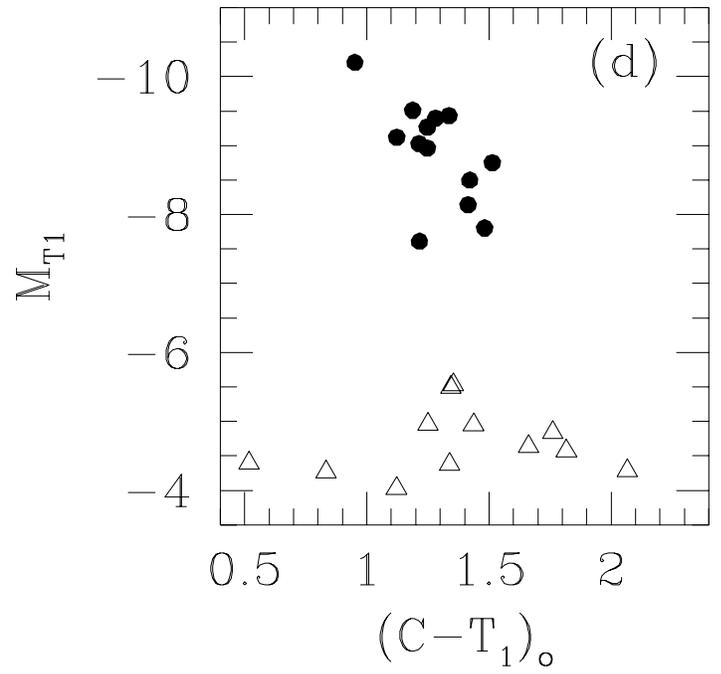
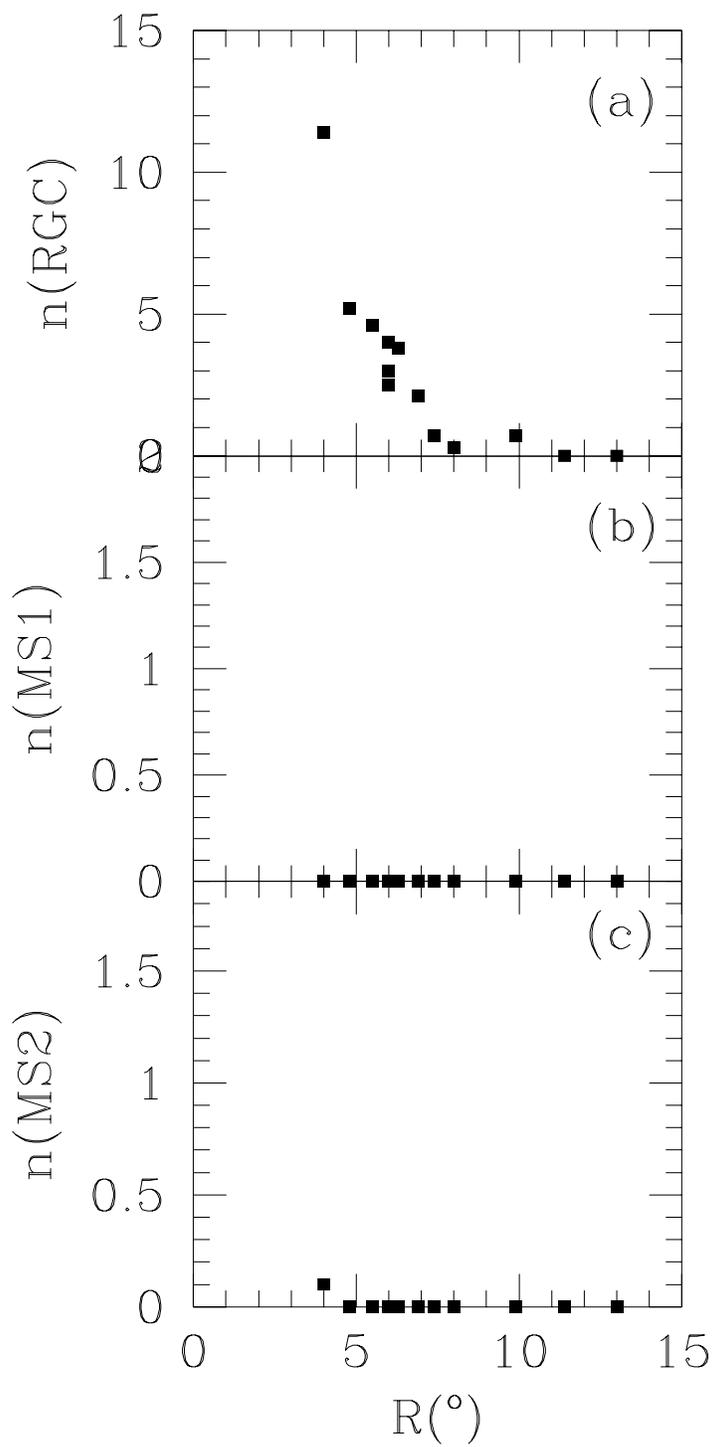
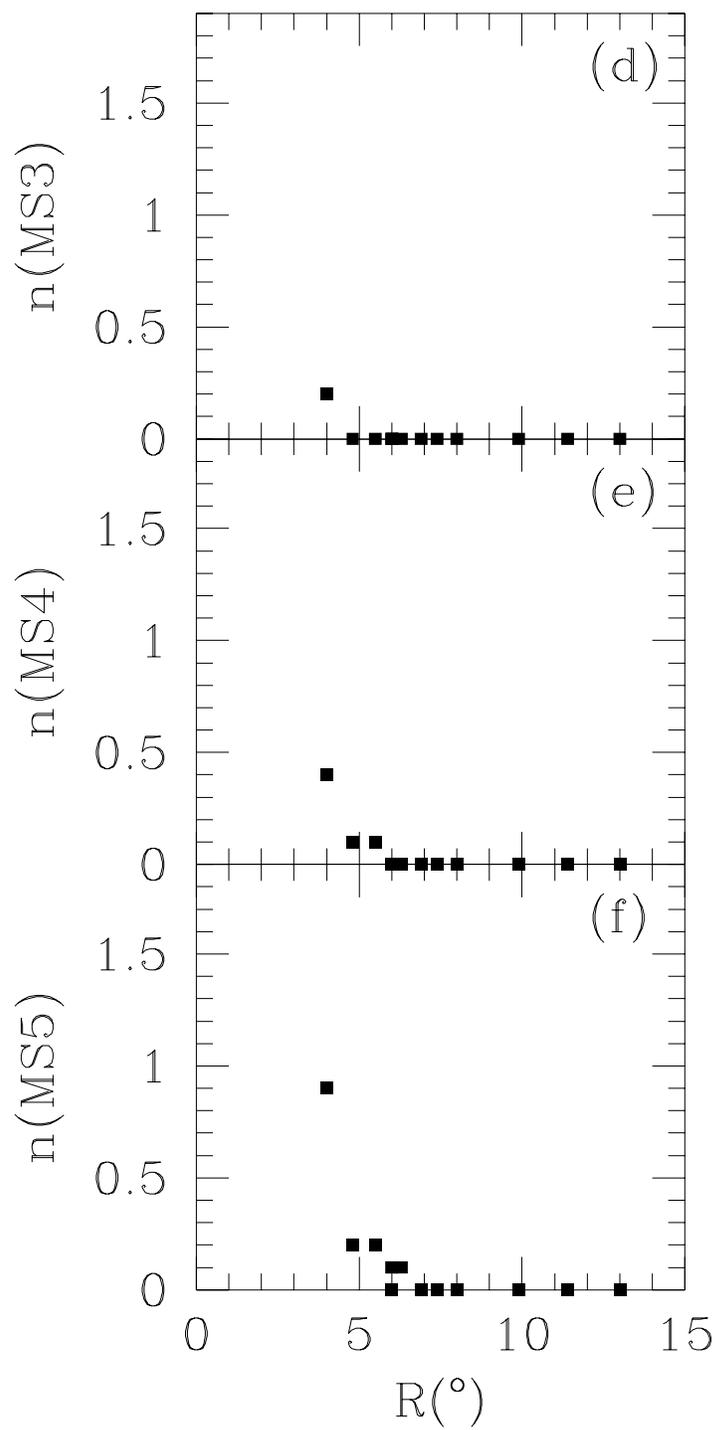
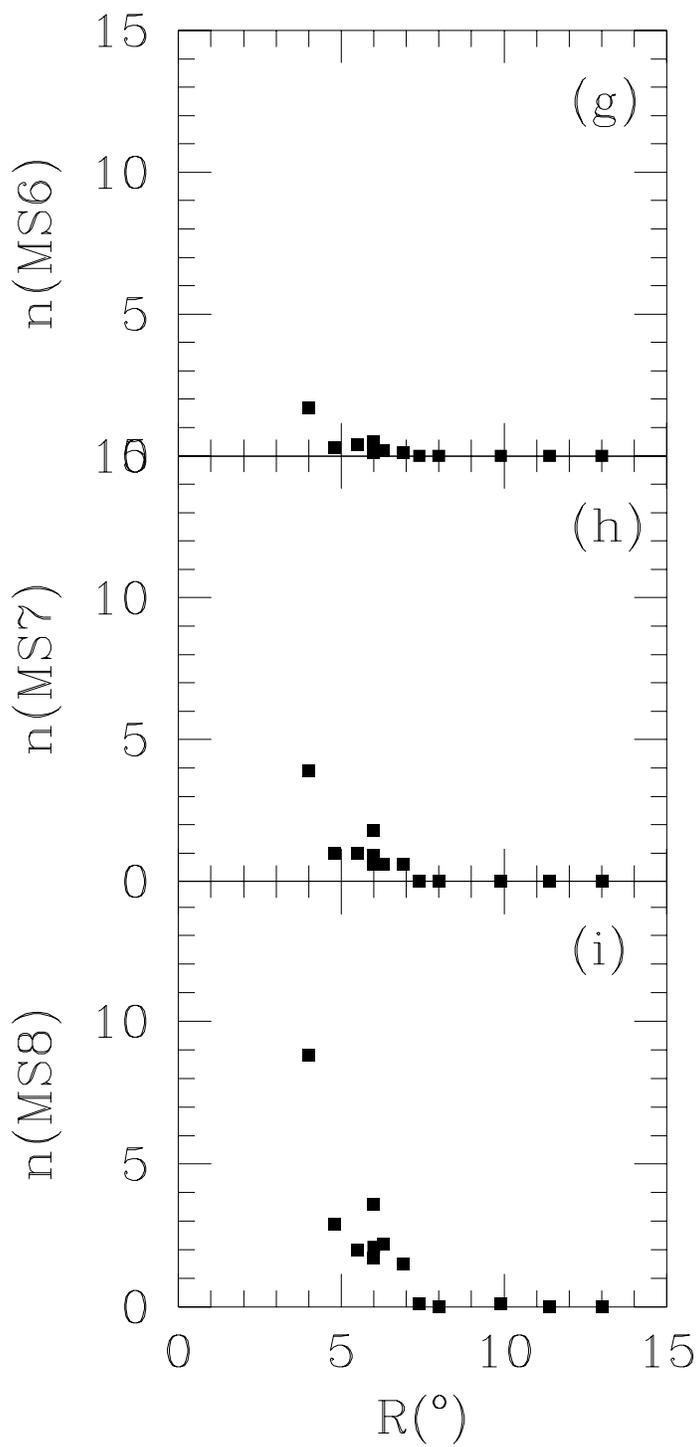
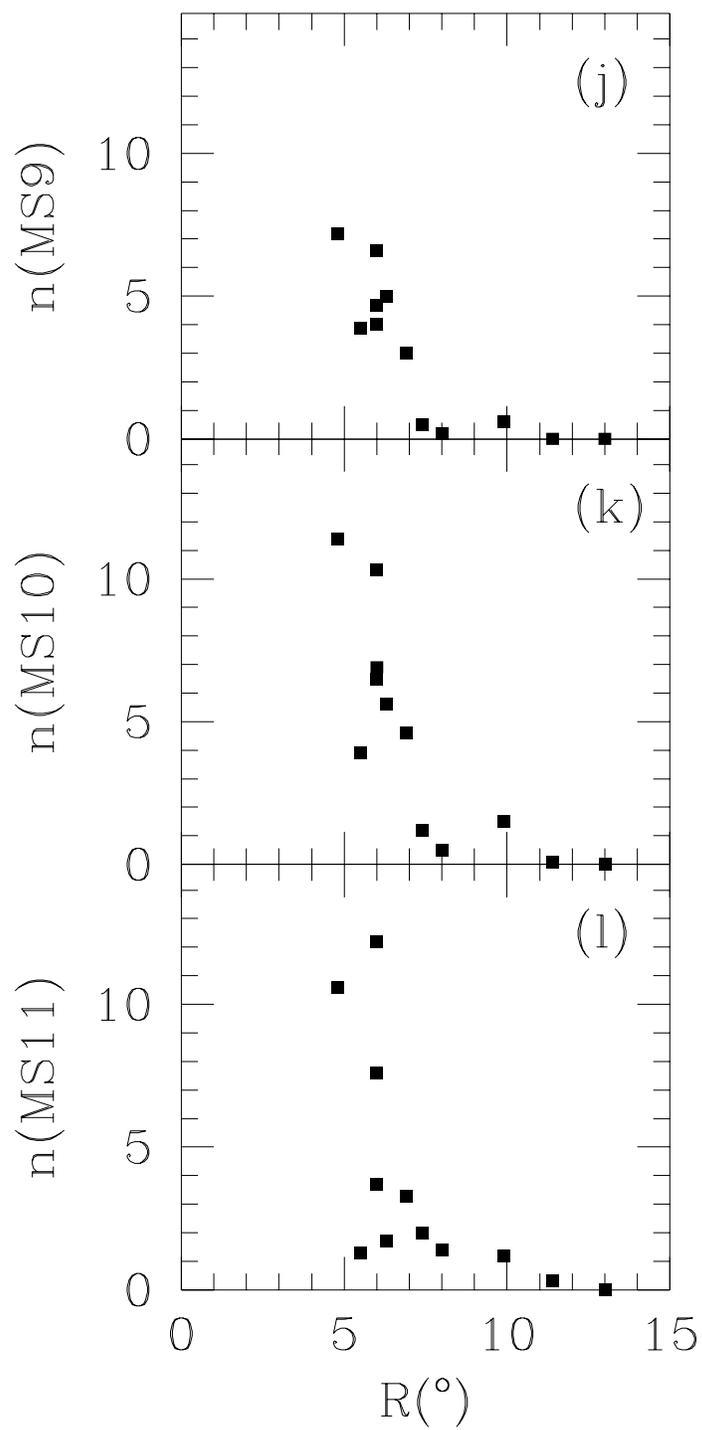
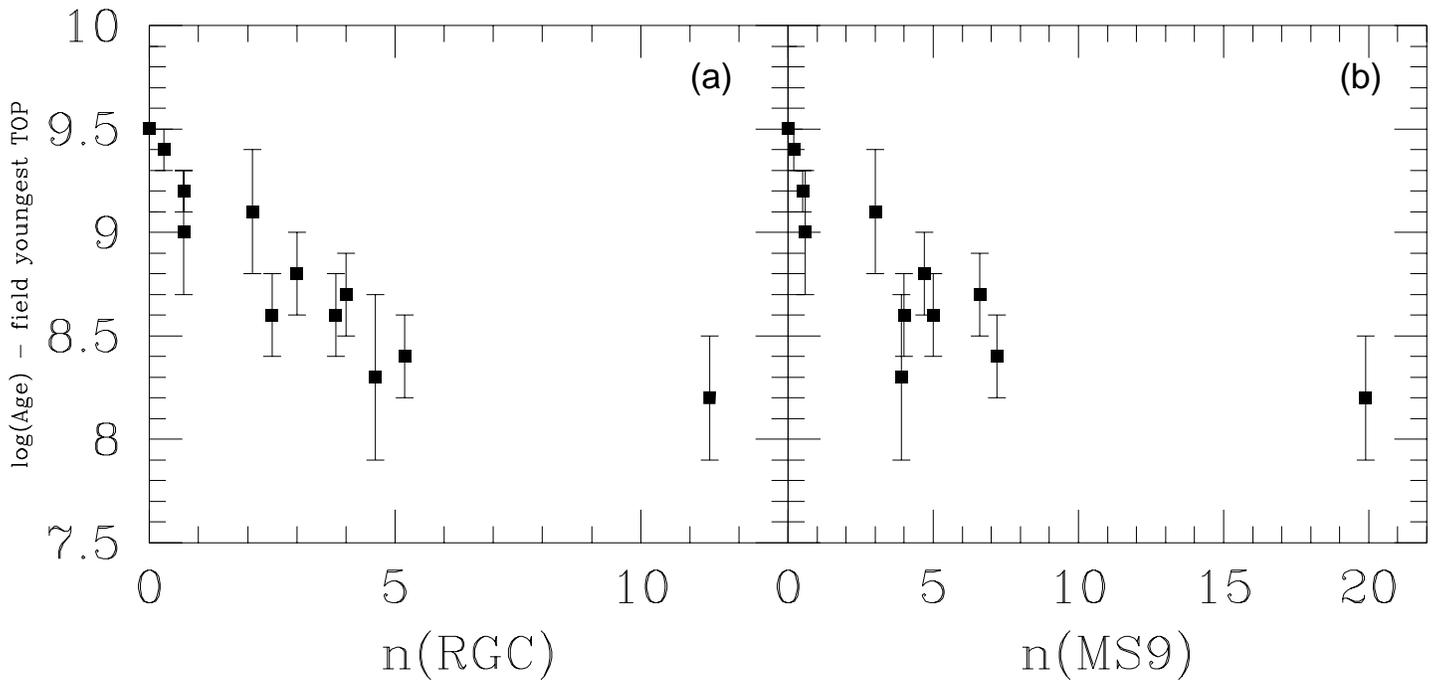
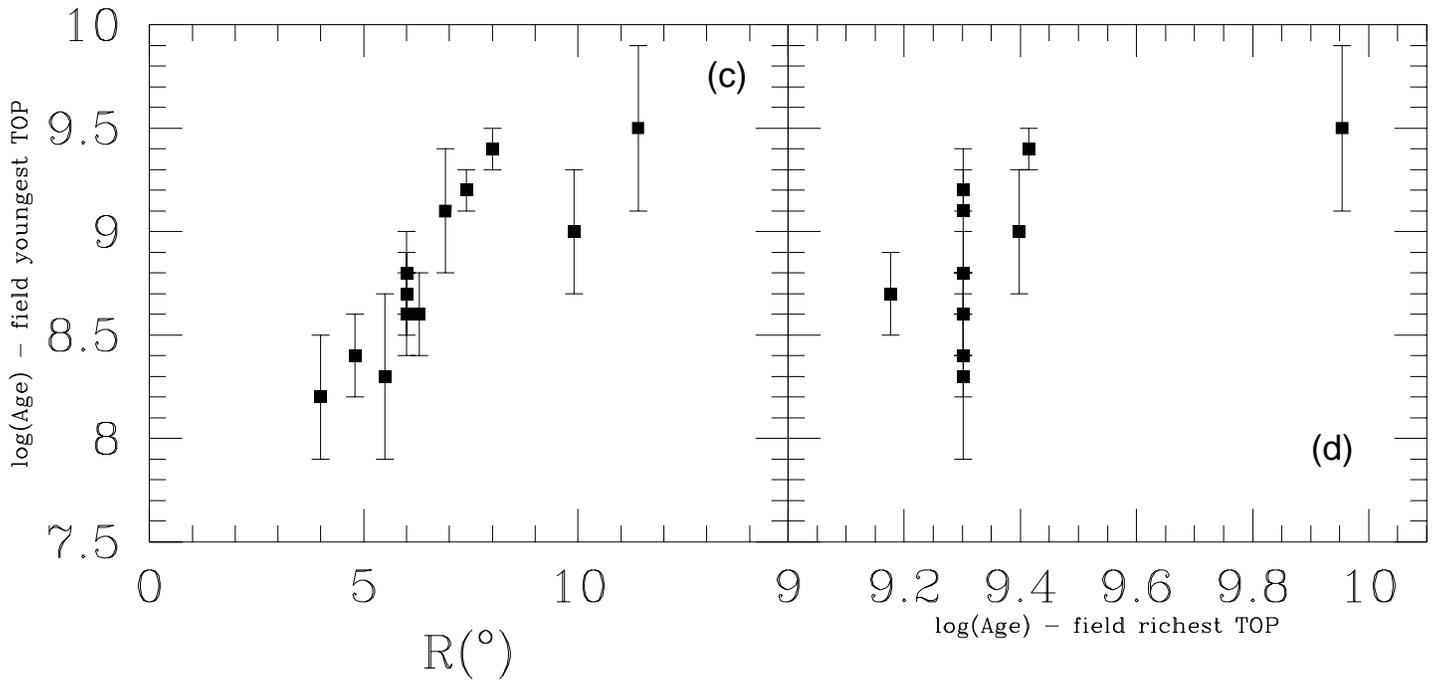
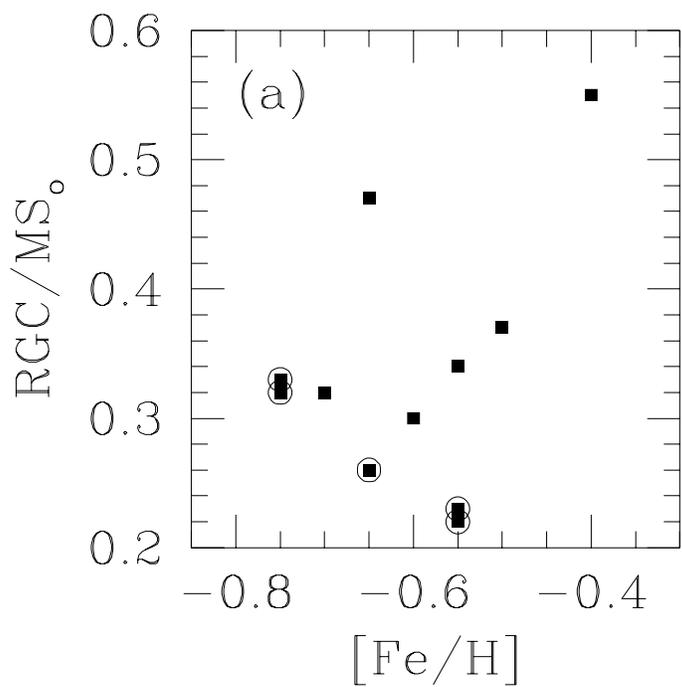
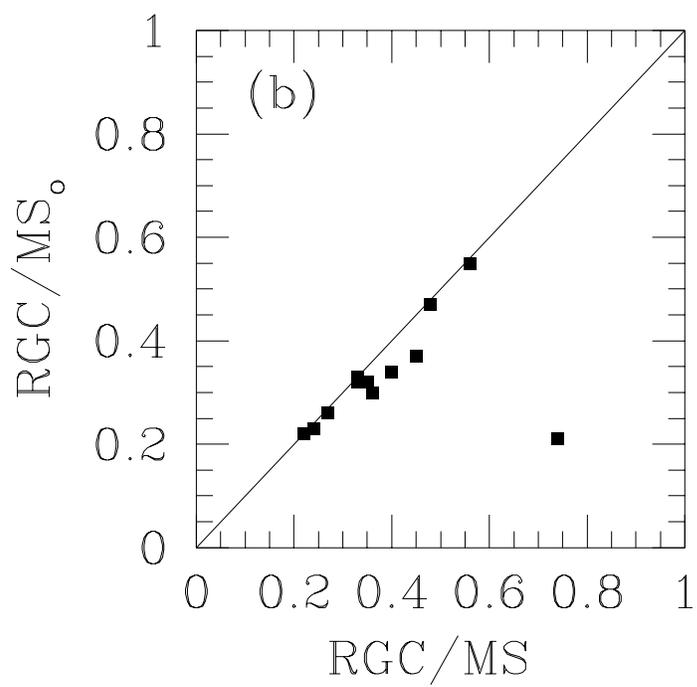
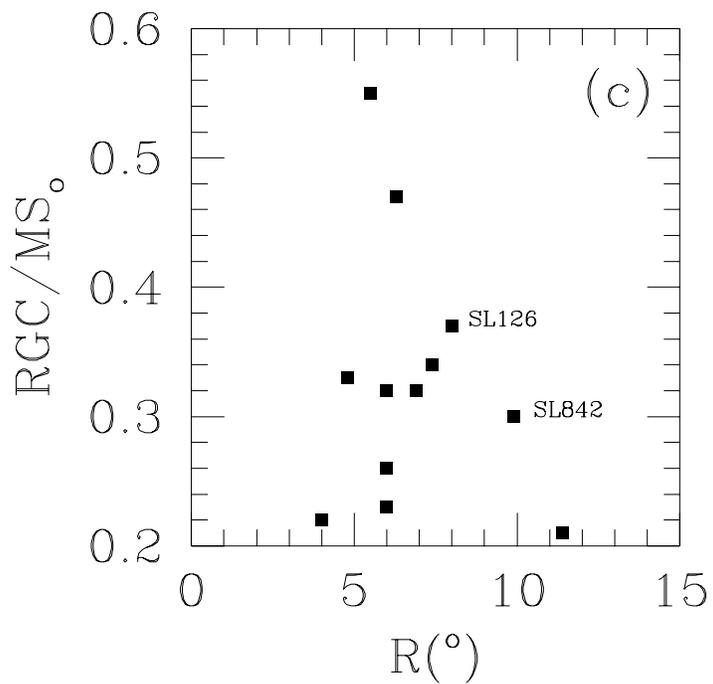
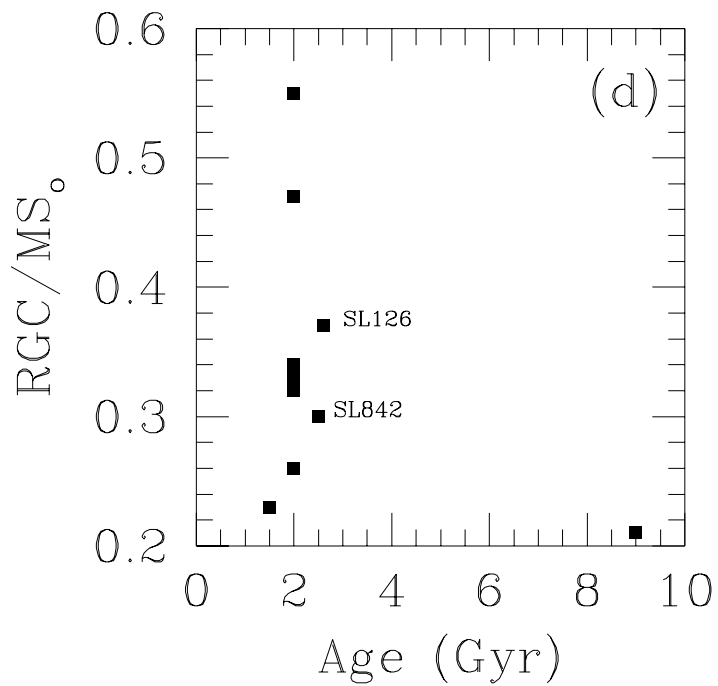
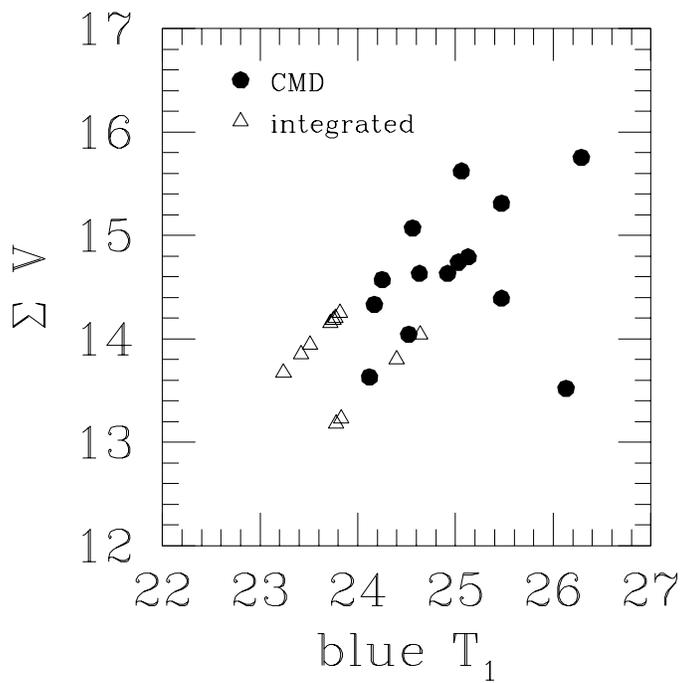
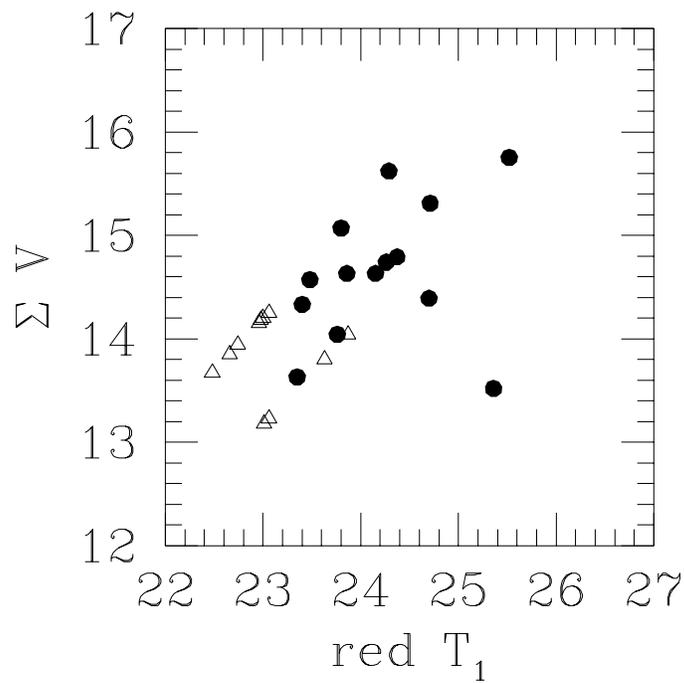
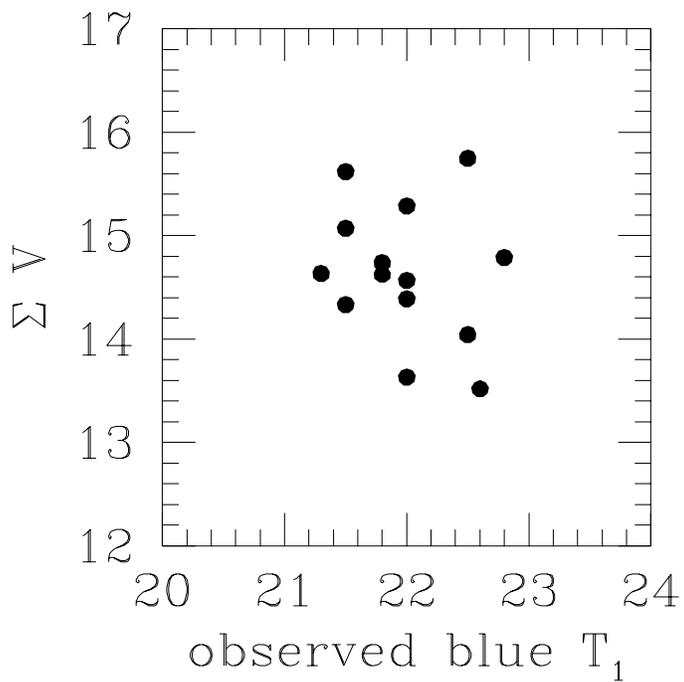
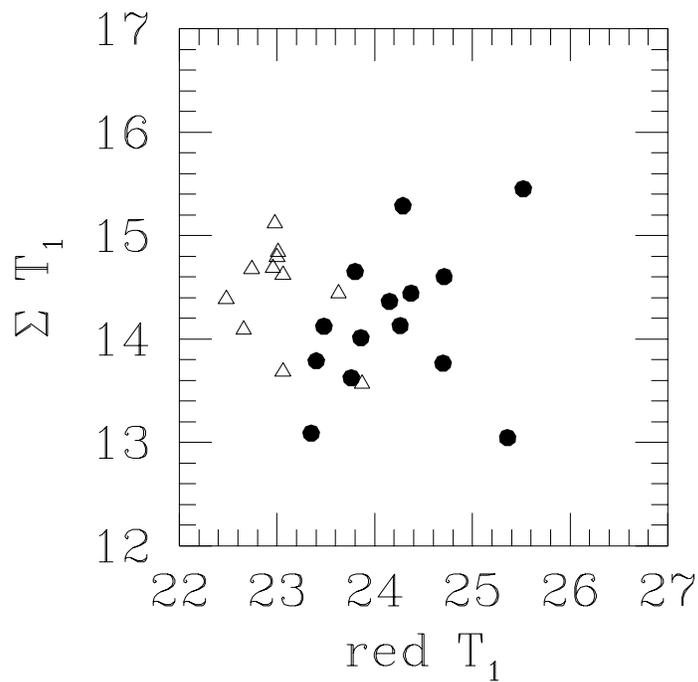
\end{document}


\scriptsize
\begin{deluxetable}{lrrlrrrrrrrr}
\tablenum{1}
\tablecolumns{12}
\tablewidth{0pt}
\tablecaption{Clusters and surrounding fields}
\tablehead{
\colhead{Cluster ID}& \colhead{$\alpha_{1950}$}&\colhead{$\delta_{1950}$}&
\colhead{l($^o$)}&\colhead{b($^o$)}&\colhead{N}&\colhead{r($^o$)}&
\colhead{R($^o$)}&\colhead{E($C-T_1$)}&\colhead{($T_1-M_{T1}$)}&
\colhead{Age}&\colhead{$[Fe/H]$} \\ 
\colhead{}& \colhead{(h  m  s)} &\colhead{($^o$  '  '')}&\colhead{}&\colhead{}&\colhead{}&\colhead{}&\colhead{}&\colhead{}&\colhead{}&\colhead{(Gyr)}&
\colhead{}  }
\startdata
SL8,LW13          & 04 38 00 &-69 07 37 &  280  &  -38.2 &   75 & 3.9&5.5&0.08& 18.60&1.8:&-0.50\\
~~~~~field        &          &          &       &        & 4355 &&&&&2.0& -0.35   \\
SL126,ESO85SC21   & 04 56 53 &-62 36 37 &  272.5&  -36.9 &  101 & 7.5&8.0&0.02& 18.53&2.2&-0.45:\\
~~~~~field        &          &          &       &        & 1437 &&&&&$>2.5$& -0.45   \\
SL262,LW146       & 05 08 52 &-62 26 27 &  272.0&  -36.0 &  104 & 7.2&7.4&0.00& 18.50&2.1&-0.55\\
~~~~~field        &          &          &       &        & 1737 &&&&&2.0& -0.50   \\
SL388,LW186       & 05 19 44 &-63 31 44 &  273.1&  -34.2 &  200 & 6.0&6.0&0.06& 18.58&2.2&-0.65\\
~~~~~field        &          &          &       &        & 6421 &&&&&2.0& -0.60   \\
IC2134,SL437,LW198\tablenotemark{a} &05 24 47&-75 29 26 &287.1&-31.7&119&6.0&6.0&0.20& 18.76&1.0&-\\
~~~~~field        &          &          &       &        & 4885 &&&&&2.0& -0.70   \\
SL451,LW206\tablenotemark{a}        &05 25 57&-75 36 33 &286  &-31.6& 68&6.1&6.1&0.20& 18.76&2.2&-0.70:\\
~~~~~field        &          &          &       &        & 4885 &&&&&2.0& - \\
SL509,LW221       & 05 29 29 &-63 41 11 &  273.2&  -33.1 &  217 & 5.9&6.0&0.06& 18.58&1.2&-0.85\\
~~~~~field        &          &          &       &        &10131 &&&&&1.5& -0.50   \\
SL769\tablenotemark{b}              &05 53 54&-70 04 44 &281  &-30.3&203&2.8&4.0&0.16& 18.71&1.8&-0.45\\
~~~~~field        &          &          &       &        &57222 &&&&&-& -0.50   \\
SL817             & 06 01 09 &-70 04 07 &   280 &  -29.7 &   66 & 3.4&4.8&0.14& 18.68&1.5&-0.50:\\
~~~~~field        &          &          &       &        &10142 &&&&&2.0& -0.70   \\
ESO121SC03\tablenotemark{c}         &06 01 27&-60 31 17 &269.5&-29.4&394&10.3&11.4&0.06& 18.58&8.5&-1.05\\
~~~~~field        &          &          &       &        &  441 &&&&&9.0& -   \\
SL842,LW399       & 06 07 53 &-62 58 40 &  272.3&  -28.8 &   57 & 8.4&9.9&0.06& 18.58&2.2&-0.60\\
~~~~~field        &          &          &       &        & 1591 &&&&&2.5& -0.55   \\
SL862,LW431       & 06 14 04 &-70 40 45 &  281.1&  -28.6 &  110 & 4.5&6.3&0.18& 18.74&1.8&-0.85\\
~~~~~field        &          &          &       &        & 4772 &&&&&2.0& -0.60   \\
OHSC33            & 06 16 33 &-73 45 58 &  284  &  -28.5 &   47 & 5.7&6.9&0.18& 18.74&1.4&-1.00\\
~~~~~field        &          &          &       &        & 4022 &&&&&2.0& -0.65   \\
OHSC37            & 07 08 01 &-69 54 10 &  280  &  -24.0 &   27 & 9.2&13.0&0.29& 18.89&2.1&-0.65\\
~~~~~field        &          &          &       &        &  559 &&&&&-& -  \\
\enddata
\tablenotetext{}{Ages, metallicities and photometric indices are from 
Bica et al. (1998).}
\tablenotetext{a}{IC2134 and SL451 are located in the same frame.} 
\tablenotetext{b}{SL769 is an inner disk cluster.} 
\tablenotetext{c}{ESO121SC03 is the oldest cluster in the sample (8.5\,Gyr).}
\end{deluxetable}

\normalsize
\begin{deluxetable}{lrr}
\tablenum{2}
\tablecolumns{3}
\tablewidth{0pt}
\tablecaption{Star Densities ($N/\Box$')}
\tablehead{
\colhead{V bin ($\pm1$)}& \colhead{OHSC\,37 Field}& \colhead{BS80 Model} }
\startdata
14              &   0.05  (10)   & 0.13$\pm$0.03  \\
16              &   0.24  (43)   & 0.36$\pm$0.09  \\
18              &   0.82 (149)   & 0.96$\pm$0.24  \\
20              &   1.30 (236)   & 1.87$\pm$0.47  \\
22              &   0.67 (121)   & 1.85$\pm$0.46  \\
Total (18$\pm$5)  &   3.07 (559)   & 5.17$\pm$1.29  \\
\enddata
\tablenotetext{}{Values parenthesized are number of stars within the
$V$ bin. Last column shows errors according to BS80.}
\end{deluxetable}

\scriptsize
\begin{deluxetable}{lrrrrrrrrrrrr}
\tablenum{3}
\tablecolumns{13}
\tablewidth{400pt}
\tablecaption{Integrated Light Properties}
\tablehead{
\colhead{Cluster}& \colhead{}&\colhead{}&\colhead{$N/\Box$'}&\colhead{}&
\colhead{}& \colhead{$M_v$}&\colhead{}&\colhead{}&\colhead{$M_{T1}$}&
\colhead{}& \colhead{}&\colhead{($C-T_1$)$_o$}  }
\startdata
SL8            &&&   53.71  &&&  -4.24  &&&  -4.83  &&&   1.76  \\  
~~~~~field     &&&   24.27  &&&  -8.98  &&&  -9.43  &&&   1.33  \\  
SL126          &&&  128.59  &&&  -2.78  &&&  -3.07  &&&   0.85  \\  
~~~~~field     &&&    7.90  &&&  -7.32  &&&  -7.80  &&&   1.48  \\  
SL262          &&&  297.93  &&&  -3.93  &&&  -4.37  &&&   1.33  \\  
~~~~~field     &&&    9.46  &&&  -7.68  &&&  -8.14  &&&   1.41  \\  
SL388          &&&  254.64  &&&  -4.55  &&&  -4.95  &&&   1.25  \\  
~~~~~annulus 1 &&&   42.02  &&&  -6.64  &&&  -7.07  &&&   1.30  \\  
~~~~~annulus 2 &&&   38.03  &&&  -7.17  &&&  -7.55  &&&   1.15  \\  
~~~~~annulus 3 &&&   33.09  &&&  -6.50  &&&  -6.86  &&&   1.04  \\  
~~~~~field     &&&   35.31  &&&  -8.61  &&&  -9.02  &&&   1.21  \\  
IC2134         &&&  151.51  &&&  -4.19  &&&  -4.39  &&&   0.51  \\  
~~~~~field     &&&   27.34  &&&  -8.86  &&&  -9.26  &&&   1.24  \\  
SL451          &&&   86.58  &&&  -4.08  &&&  -4.63  &&&   1.66  \\  
~~~~~field     &&&   27.34  &&&  -8.86  &&&  -9.26  &&&   1.24  \\  
SL509          &&&  276.29  &&&  -4.96  &&&  -5.49  &&&   1.34  \\  
~~~~~annulus 1 &&&   58.77  &&&  -7.11  &&&  -7.49  &&&   1.11  \\  
~~~~~annulus 2 &&&   56.76  &&&  -6.88  &&&  -7.32  &&&   1.22  \\  
~~~~~annulus 3 &&&   60.24  &&&  -6.85  &&&  -7.25  &&&   1.10  \\  
~~~~~field     &&&   55.72  &&&  -8.72  &&&  -9.12  &&&   1.12  \\  
SL769          &&&  258.46  &&&  -3.96  &&&  -4.26  &&&   0.83  \\  
~~~~~annulus 1 &&&  333.57  &&&  -7.90  &&&  -8.30  &&&   1.00  \\  
~~~~~annulus 2 &&&  330.73  &&&  -7.95  &&&  -8.32  &&&   0.93  \\  
~~~~~annulus 3 &&&  326.01  &&&  -7.97  &&&  -8.32  &&&   0.90  \\  
~~~~~annulus 4 &&&  312.28  &&&  -8.06  &&&  -8.44  &&&   0.97  \\  
~~~~~field     &&&  314.72  &&&  -9.83  &&& -10.20  &&&   0.95  \\  
SL817          &&&  189.07  &&&  -3.65  &&&  -4.03  &&&   1.12  \\  
~~~~~field     &&&   55.25  &&&  -9.10  &&&  -9.51  &&&   1.18  \\  
ESO121SC03     &&&   70.54  &&&  -5.07  &&&  -5.53  &&&   1.35  \\  
~~~~~field     &&&    2.71  &&&  -7.21  &&&  -7.61  &&&   1.21  \\  
SL842          &&&  113.39  &&&  -3.96  &&&  -4.56  &&&   1.81  \\  
~~~~~field     &&&    8.69  &&&  -8.27  &&&  -8.75  &&&   1.51  \\  
SL862          &&&  218.83  &&&  -4.45  &&&  -4.95  &&&   1.43  \\  
~~~~~field     &&&   26.08  &&&  -8.97  &&&  -9.39  &&&   1.28  \\  
OHSC33         &&&  134.64  &&&  -3.16  &&&  -3.45  &&&   0.78  \\  
~~~~~field     &&&   21.91  &&&  -8.55  &&&  -8.96  &&&   1.24  \\  
OHSC37         &&&   34.37  &&&  -3.67  &&&  -4.28  &&&   2.06  \\  
~~~~~field     &&&    3.07  &&&  -8.06  &&&  -8.49  &&&   1.42  \\  
\enddata
\tablenotetext{}{$field$ entries are related to the CCD area minus the
area corresponding to twice the cluster radius; $annulus$ entries are
related to ring areas corresponding to the radius limit $150<r(pix)<500$
(annulus 1), $500<r(pix)<691$ (2), $691<r(pix)<840$ (3) and
$840<r(pix)<966$ (4).} 
\end{deluxetable}

\scriptsize

\begin{deluxetable}{lrrrrrr}
\tablenum{4}
\tablecolumns{7}
\tablewidth{0pt}
\tablecaption{Threshold $T_1$ Magnitude For Crowding Evaluation}
\tablehead{
\colhead{Cluster}& \colhead{$V$}&\colhead{diaph.('')}&\colhead{$l_T$}&
\colhead{$T_1$ blue}&\colhead{$T_1$ red}& \colhead{$T_1$ obs. blue} }
\startdata
SL8      & 14.39  & 80.00  &  0.22  & 25.47  & 24.70 &22.00\\ 
      & 13.80  & 61.00  &  0.59  & 24.40  & 23.63 &\\ 
SL126     & 15.75  & 60.00  &  0.10  & 26.29  & 25.52 &22.50\\ 
     & -  & -  &  -  &-  & - &\\ 
SL262     & 14.57  & 40.00  &  0.61  & 24.25  & 23.48 &22.00\\ 
     & 14.25  & 38.00  &  0.90  & 23.82  & 23.06 &\\ 
SL388     & 14.04  & 60.00  &  0.51  & 24.52  & 23.76 & 22.50\\ 
     & 13.85  & 38.00  &  1.42  & 23.42  & 22.66 &\\ 
IC2134     & 14.63  & 60.00  &  0.44  & 24.92  & 24.15 &21.80\\ 
     & 13.94  & 38.00  &  1.60  & 23.51  & 22.74 &\\ 
SL451     & 14.74  & 60.00  &  0.39  & 25.03  & 24.26 &21.80\\ 
     & 14.20  & 38.00  &  1.26  & 23.77  & 23.00 &\\ 
SL509     & 13.63  & 60.00  &  0.75  & 24.12  & 23.35 &22.00\\ 
     & 13.23  & 61.00  &  0.97  & 23.83  & 23.06 &\\ 
SL769     & 14.79  & 60.00  &  0.34  & 25.13  & 24.37 & 22.80\\ 
     & 13.18  & 61.00  &  1.18  & 23.78  & 23.01 &\\ 
SL817     & 15.07  & 40.00  &  0.55  & 24.56  & 23.80 &21.50\\ 
     & 14.18  & 38.00  &  1.18  & 23.75  & 22.98 &\\ 
ESO121SC03   & 13.52  &160.00  &  0.12  & 26.13  & 25.36 &22.60\\ 
   & 14.04  & 61.00  &  0.46  & 24.64  & 23.87 &\\ 
SL842     & 14.63  & 48.00  &  0.47  & 24.63  & 23.86 &21.30\\ 
     & 14.15  & 38.00  &  1.08  & 23.72  & 22.96 &\\ 
SL862     & 14.33  & 48.00  &  0.85  & 24.17  & 23.40 &21.50\\ 
     & 13.67  & 38.00  &  1.99  & 23.24  & 22.48 &\\ 
OHSC33    & 15.62  & 40.00  &  0.37  & 25.06  & 24.29 &21.50\\ 
    & -  & -  &  -  &-  &- &\\ 
OHSC37    & 15.31  & 60.00  &  0.30  & 25.47  & 24.71 &22.00\\ 
    & -  & -  & -   &-  &- &\\ 
\enddata
\tablenotetext{}{Column (2) gives the integrated $V$ magnitude observed
by Bica et al. (1996) with the diaphragm used in column (3); Column
(4) is the corresponding bolometric luminosity per pixel (see text for
details); columns (5), (6) and (7) show the threshold magnitude where 
crowding starts to become significant for a blue star (($C-T_1$)$_o$=0),
a red one ($C-T_1$)$_o$=3) and for the observed faintest blue stars sampled in
the CMDs}
\end{deluxetable}